\documentclass[prd,twocolumn,nofootinbib,amsmath,amssymb]{revtex4-1}
\usepackage{graphicx}
\usepackage{color}
\usepackage{amssymb}
\usepackage{subfigure}
\usepackage{natbib}
\usepackage{color}
\usepackage[colorlinks]{hyperref}
\usepackage{amsmath}
\usepackage{mathrsfs}

\def\gsim{ \lower .75ex \hbox{$\sim$} \llap{\raise .27ex \hbox{$>$}} }

\def\lsim{ \lower .75ex \hbox{$\sim$} \llap{\raise .27ex \hbox{$<$}} }

\def\IZ{\relax\ifmmode\mathchoice
{\hbox{\cmss Z\kern-.4em Z}}{\hbox{\cmss Z\kern-.4em Z}}
{\lower.9pt\hbox{\cmsss Z\kern-.4em Z}} {\lower1.2pt\hbox{\cmsss
Z\kern-.4em Z}}\else{\cmss Z\kern-.4em Z}\fi}
\def\IR{\relax{\rm I\kern-.18em R}}

\def\one{{\hbox{ 1\kern-.8mm l}}}

\newlength{\bredde}
\def\slash#1{\settowidth{\bredde}{$#1$}\ifmmode\,\raisebox{.15ex}{/}
\hspace*{-\bredde} #1\else$\,\raisebox{.15ex}{/}\hspace*{-\bredde}
#1$\fi}

\newsavebox{\zzzbar}
\sbox{\zzzbar}
  {\setlength{\unitlength}{0.9em}
  \begin{picture}(0.6,0.7)
  \thinlines
  \put(0,0){\line(1,0){0.6}}
  \put(0,0.75){\line(1,0){0.575}}
  \multiput(0,0)(0.0125,0.025){30}{\rule{0.3pt}{0.3pt}}
  \multiput(0.2,0)(0.0125,0.025){30}{\rule{0.3pt}{0.3pt}}
  \put(0,0.75){\line(0,-1){0.15}}
  \put(0.015,0.75){\line(0,-1){0.1}}
  \put(0.03,0.75){\line(0,-1){0.075}}
  \put(0.045,0.75){\line(0,-1){0.05}}
  \put(0.05,0.75){\line(0,-1){0.025}}
  \put(0.6,0){\line(0,1){0.15}}
  \put(0.585,0){\line(0,1){0.1}}
  \put(0.57,0){\line(0,1){0.075}}
  \put(0.555,0){\line(0,1){0.05}}
  \put(0.55,0){\line(0,1){0.025}}
  \end{picture}}

\newcommand{\ena}{\end{eqnarray}}
\newcommand{\beqa}{\begin{eqnarray}}
\newcommand{\eeqa}{\end{eqnarray}}
\newcommand{\bea}{\begin{eqnarray}}
\newcommand{\eea}{\end{eqnarray}}

\newcommand{\be}{\begin{equation}}
\newcommand{\ee}{\end{equation}}

\def\ben{\begin{equation}}
\def\een{\end{equation}}
\def\half{{1 \over 2}}
\def\bea{\begin{eqnarray}}
\def\eea{\end{eqnarray}}
\def\be{\begin{equation}}
\def\ee{\end{equation}}
\def\beq{\begin{eqnarray}}
\def\eeq{\end{eqnarray}}
\def\ba{\begin{eqnarray}}
\def\ea{\end{eqnarray}}

\input amssym.def
\input amssym.tex
\begin{document}

\title{Particle on the Innermost Stable Circular Orbit of a Rapidly Spinning Black Hole}

\author{Samuel E. Gralla}
\author{Achilleas P. Porfyriadis}
\affiliation{ Center for the Fundamental Laws of Nature, Harvard University, Cambridge, MA 02138, USA }
\author{Niels Warburton}
\affiliation{MIT Kavli Institute for Astrophysics and Space Research, Massachusetts Institute of Technology, Cambridge, MA 02139, USA}

\begin{abstract}
We compute the radiation emitted by a particle on the innermost stable circular orbit of a rapidly spinning black hole both (a) analytically, working to leading order in the deviation from extremality and (b) numerically, with a new high-precision Teukolsky code. We find excellent agreement between the two methods.  We confirm previous estimates of the overall scaling of the power radiated, but show that there are also small oscillations all the way to extremality.  Furthermore, we reveal an intricate mode-by-mode structure in the flux to infinity, with only certain modes having the dominant scaling.  The scaling of each mode is controlled by its conformal weight, a quantity that arises naturally in the representation theory of the enhanced near-horizon symmetry group.  We find relationships to previous work on particles orbiting in precisely extreme Kerr, including detailed agreement of quantities computed here with conformal field theory calculations performed in the context of the Kerr/CFT correspondence.
\end{abstract}

\maketitle

\section{Introduction}

Unlike its simple Newtonian counter-part, the general relativistic two-body problem is a sprawling collection of different regimes, each with its own special techniques, where it becomes possible to precisely define and solve the problem.  In recent years this two-body landscape has been explored in impressive detail, driven primarily by the need for accurate theoretical models of gravitational-wave sources.  Well-separated masses are treated with high-order post-Newtonian expansions, large mass-ratio cases are treated with point particle perturbation theory, and close orbits of comparable mass systems are handled with numerical simulations.  Non-trivial checks in overlapping domains of validity \cite{letiec2014} give confidence that these diverse efforts are converging towards what could be called a complete solution of the relativistic two-body problem.

One corner just beginning to be filled in \cite{porfyriadis-strominger2014,hadar-porfyriadis-strominger2014,hadar-porfyriadis-strominger2015} is that of a particle orbiting in the near-horizon region of a near-extreme Kerr black hole.  From a theoretical perspective, this is one of the most interesting regimes since it enjoys an enhanced isometry group as well as an infinite-dimensional asymptotic symmetry group \cite{bardeen-horowitz1999,kerrCFT}.  For practical purposes, calculations at extremes of parameter space can provide useful calibration points for approximation schemes, such as the effective one-body formalism \cite{buonanno-damour1999}, aiming to be uniform over parameter space.  Finally, thought experiments showing naive violation of the cosmic censorship conjecture by throwing particles into a near-extreme black hole \cite{hubeny1999,jacobson-sotiriou2009} provide additional motivation to study near-horizon, near-extreme orbits.

In this article we compute the radiation from a particle on the innermost stable circular orbit (ISCO) of a rapidly spinning Kerr black hole.  This radiation plays an important role in the transition from inspiral to plunge \cite{ori-thorne2000,kesden2011} and also informs studies of the validity of the cosmic censorship conjecture \cite{Barausse:2010ka,barausse-cardoso-khanna2011,colleoni-barack2015}.  Previous work \cite{chrzanowski1976,kesden2011,colleoni-barack2015} has estimated the scaling near extremality to be $p=2/3$, where the total energy radiated per unit time is expressed as
\begin{equation}\label{power}
\dot{\mathcal{E}} = C \epsilon^p, \qquad \epsilon \equiv \sqrt{1-a^2/M^2},
\end{equation}
with $M$ and $Ma$ the mass and spin of the black hole.

Our calculations confirm $p=2/3$ but also reveal some interesting details.  First, the coefficient $C$ is not a constant, and instead exhibits oscillations in $\epsilon$ about its mean value.  Second, there is an intricate structure in the $\ell,m$ angular modes of the radiation.  While all modes have $p=2/3$ for the flux down the horizon, the same is not true for the flux at infinity.  Instead, the exponent for the power at infinity is given by
\begin{equation}
p_{\infty} = \frac{4}{3} \textrm{Re}[h], \qquad
\end{equation}
where $h$ is the \textit{conformal weight} of the mode, given in terms of the angular eigenvalues $\{K,m\}$ (spheroidal and azimuthal) by
\begin{equation}\label{h}
 h \equiv \frac{1}{2}+\sqrt{K-2 m^2+\frac{1}{4}}.
\end{equation}
The notion of a conformal weight arises in the representation theory of the near-horizon symmetry group (App.~\ref{app:symmetry}) and is a key entry in the Kerr/CFT dictionary.  The weight $h$ should be thought of as fundamental, with the formula \eqref{h} depending on conventional choices like the definition of $K$.  The appearance of the conformal weight in the radiation at infinity can be interpreted as a far-field signature of the near-field symmetry enhancement.

\begin{figure}
\centering\hskip-5mm
\includegraphics[width=8.5cm]{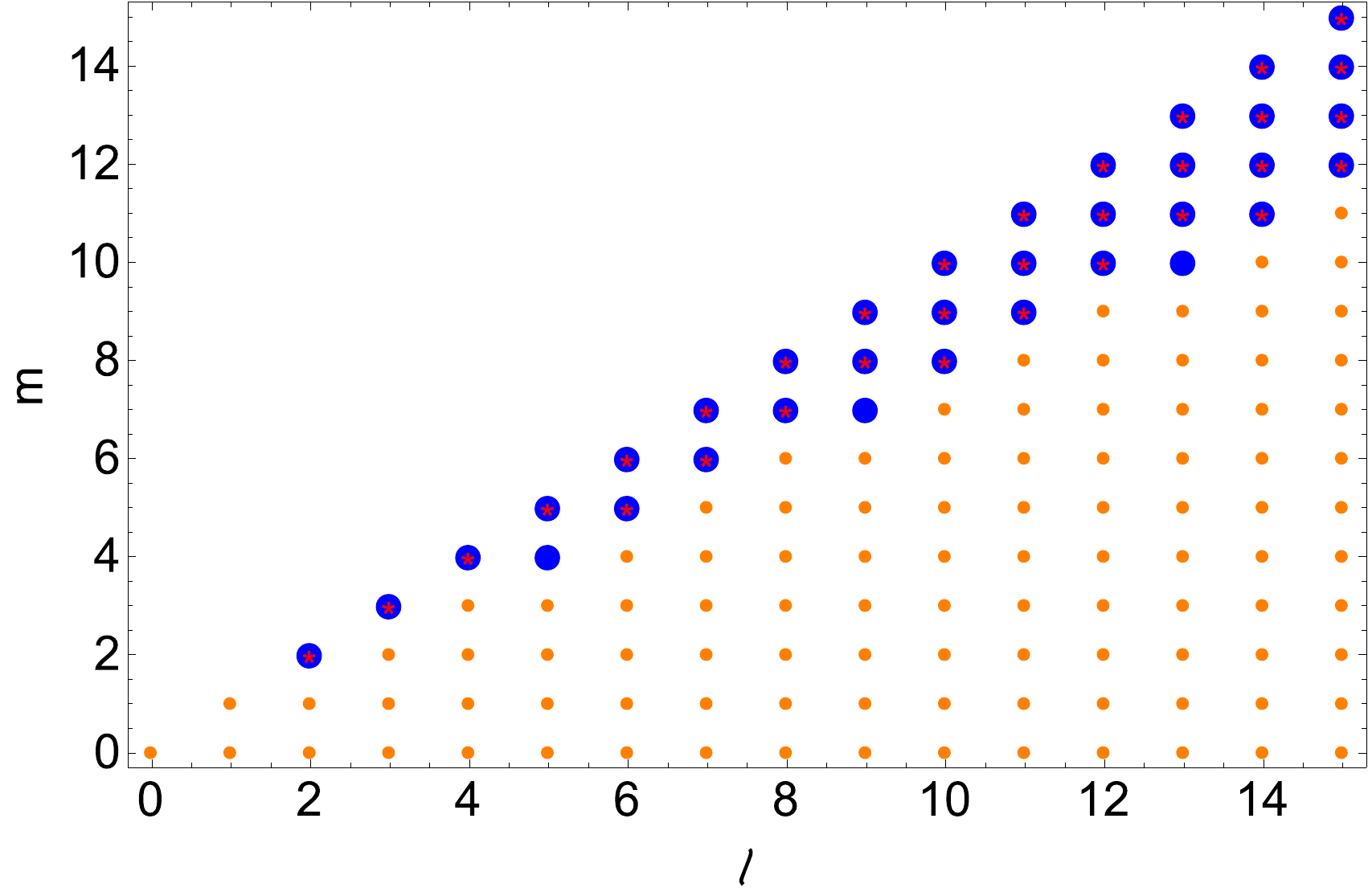}
\caption{Diagram indicating the scaling \eqref{power} of energy radiated to infinity for each mode.  Blue dots indicate the dominant scaling $p=2/3$ in the gravitational case, while red stars indicate the dominant scaling $p=2/3$ in the scalar case.  Yellow dots indicate subdominant scaling $p>2/3$.  The flux down the horizon always has dominant scaling $p=2/3$.}
\label{fig:modes}
\end{figure}

The conformal weight controls the character of each mode.  Modes with complex weight ($K-2 m^2+1/4 < 0$) have ${\rm Re}[h]=1/2$ and hence the dominant scaling $p=2/3$, while modes with real weight ($K-2 m^2+1/4 > 0$) have ${\rm Re}[h]>1/2$ and hence subdominant scaling $p>2/3$.\footnote{An analogous mode structure was previously observed in the study of near-extremal quasi-normal modes \cite{yang-etalOne2013,yang-etal2013}.}  Only the dominant modes display the oscillations in the prefactor $C$.  At each $\ell$, modes with higher values of $|m|$ are dominant (Fig.~\ref{fig:modes}).  The transition is increasingly sharp as extremality is approached, and (e.g.) already at $\epsilon=0.1$ ($a=0.995M$), the 2-2 mode dominates the 2-1 mode by four orders of magnitude (Fig.~\ref{fig:infFlux}).  An observation of a huge difference in power between the 2-2 and 2-1 modes would signal the presence of a near-extreme black hole.\footnote{A detector at a fixed position cannot probe angular dependence, but for a circular orbit the difference between $m=1$ and $m=2$ is visible in the associated time-dependence $e^{i m \Omega t}$.}

We compute the radiation both analytically (to leading order in $\epsilon$) and numerically (at small, finite $\epsilon$).  Comparing the results, we begin seeing agreement (to about 10\%) at $\epsilon=0.01$ ($a=0.99995M$) and we achieve eight digits of accuracy by the time we reach $\epsilon=10^{-13}$, the smallest value we simulate.  Historically, the near-extremal region of parameter space has been difficult to access numerically.   Our new codes mark a substantial improvement over previous work and can accurately calculate the radiated fluxes for spins as high as $a=0.999999999999999999999999995M$.

Our analytic solution of the Teukolsky equation uses the method of matched asymptotic expansions, a technique used in \cite{Starobisky,starobisky-churilov,teukolsky-press1974} and many times since.  Our consideration of a particle on the ISCO complicates matters because this orbit is in a sense intermediate between the near-horizon and far regions (Fig.~\ref{fig:throats}).  The proper way to think of the extremal ISCO has been the subject of some discussion over the years, and our calculations afford an opportunity to chime in. The fate of the ISCO is discussed in Sec.~\ref{sec:overview} and App.~\ref{app:limits}.

Previous work involving one of us \cite{porfyriadis-strominger2014} considered the physically distinct problem of a particle on a circular orbit in the near-horizon region of an exactly extremal Kerr black hole, working to leading order in the deviation from the horizon.  After performing the calculation in the case of a scalar charge in this paper, we find that the power radiated is \textit{identical} to that of \cite{porfyriadis-strominger2014} with parameters identified in the natural way.  The agreement is not completely surprising since the geometry in the vicinity of the near-extremal ISCO is the same as the near-horizon geometry of exactly extremal Kerr  (the ``NHEK'' geometry, App.~\ref{app:intermediate}).  On the other hand, the agreement is highly nontrivial since the near-extremal Kerr throat contains an entire near-horizon region with a curved geometry, which is absent in extremal Kerr.  (This region is the bottom section of Fig.~\ref{fig:throats} and is described by the ``near-NHEK'' metric, App.~\ref{near-NHEK}.)

One can expect the analogous agreement to hold in the gravitational case. We therefore do not repeat the detailed calculation of the scalar case but instead rely on the gravitational results of \cite{porfyriadis-strominger2014}.\footnote{Only the flux at infinity was presented in \cite{porfyriadis-strominger2014}.  We compute the horizon flux using expressions given therein.} Identifying the two problems in the same manner as before produces analytic expressions for the power radiated by a particle on the ISCO.  We confirm these expressions numerically.  We have not identified the precise reason for the agreement (in this particular observable) between the two different problems, but we think it is a manifestation of the action of the infinite-dimensional conformal group, which can relate extremal to near-extremal physics \cite{hadar-porfyriadis-strominger2014,hadar-porfyriadis-strominger2015}. 

In Sec.~\ref{sec:overview} we give an overview of near-extremal physics and establish notation.  In Sec.~\ref{sec:scalar} we perform the analytic calculation in the scalar case.  In Sec.~\ref{sec:gravity} we present analytic results for the gravitational case.  In Sec.~\ref{sec:numerical} we present the new numerical codes and compare the results with the analytic expressions.  An appendix reviews near-horizon limits, placing our computation in the context of this rich structure.  Our metric has signature $-+++$ and we use units with $G=c=1$.

\section{Near-Extremal Physics}\label{sec:overview}

The non-extremal Kerr black hole is invariantly characterized by two parameters $a$ and $M$ satisfying $M>0$ and $a<M$.  We will work with $M>0$ and $\epsilon>0$, where $\epsilon$ is the near-extremality parameter defined in Eq.~\eqref{power}.  It is also useful to introduce $r_\pm=M\pm\sqrt{M^2-a^2}$, the Boyer-Lindquist (BL) coordinate radii of the horizons, and the (outer) horizon angular frequency $\Omega_H=a/(r_+^2+a^2)$.  We restrict attention to $r>r_+$, which we call the Kerr exterior.

One may now ask the question: ``What is the extremal ($\epsilon \rightarrow 0$) limit of the Kerr exterior?''  Fixing Boyer-Lindquist (BL) coordinates, one obtains the spacetime conventionally called extreme Kerr.  On the other hand, fixing alternative coordinates adapted to the near-horizon region (App.~\ref{app:near-horizon}) gives a different spacetime, normally called NHEK (for near-horizon extremal Kerr).  There is yet a \textit{third} limit adapted to the ISCO, which gives a different patch of the maximally extended NHEK spacetime (App.~\ref{app:intermediate}).  The first limit leaves asymptotic infinity intact but replaces the non-degenerate horizon by a degenerate one.  The second and third limits replace asymptotic null infinity with a timelike boundary.  The answer to the question is thus ``not enough information''.  There are multiple limits and none is preferred on any fundamental grounds.

The existence of the various limits is a signal that near-extremal physics falls into the class of what are generally called \textit{singular perturbation} problems.  In our ISCO calculation, the singular nature appears as the impossibility of imposing all the boundary conditions of the differential equation in a single small-$\epsilon$ approximation.  Instead we must make a far-zone approximation where we can satisfy the far boundary conditions (no incoming radiation from past null infinity), a near-zone approximation where we can satisfy the near boundary conditions (no incoming radiation from the past horizon), and match the two in their region of overlap.

\subsection{Circular Orbits and the ISCO}\label{sec:ISCO}

\begin{figure}
\includegraphics[width=0.5\textwidth]{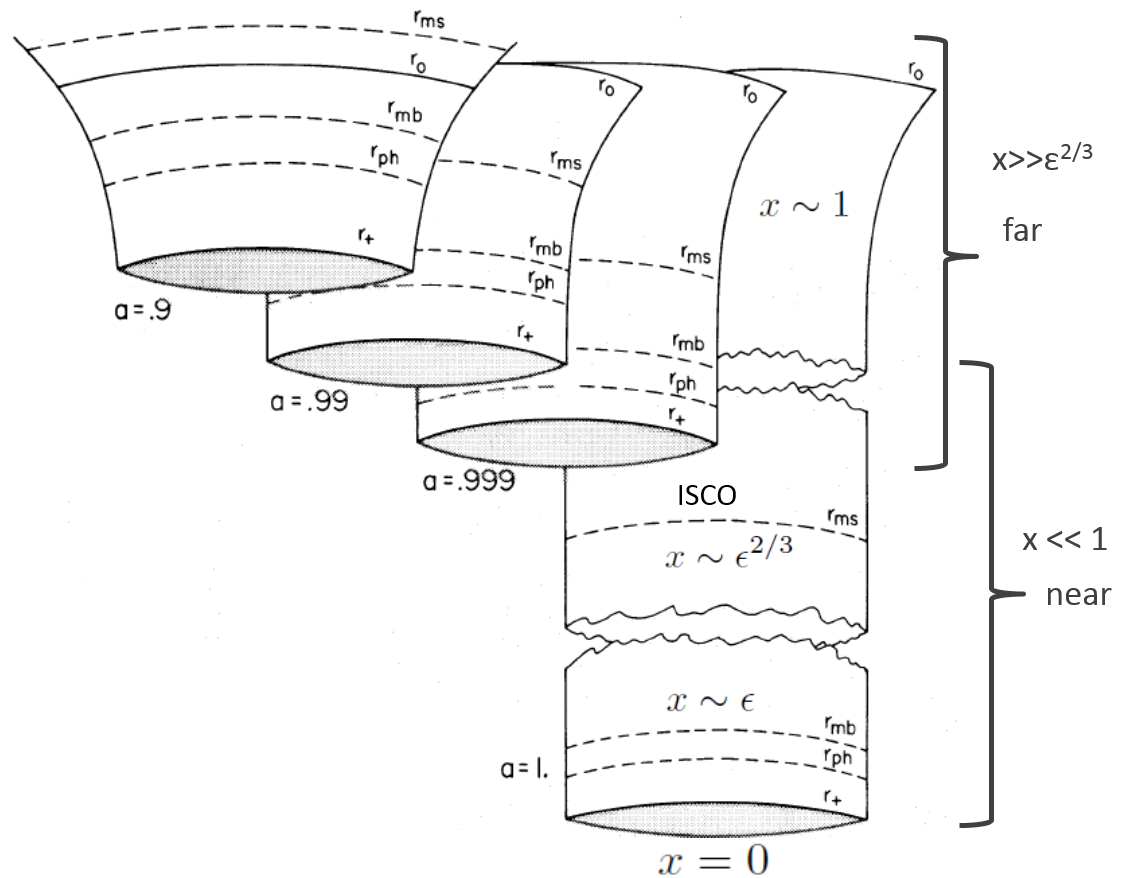}
\caption{The well-known diagram of \cite{bardeen-press-teukolsky1972} overlaid with corresponding regions of the dimensionless coordinate $x$ that we consider.  The dashed lines illustrate the BL radii of the ISCO $r_{\rm ms}$, the marginally bound orbit $r_{\rm mb}$, and the photon orbit $r_{\rm ph}$. Also shown are the horizon $r_{+}$ and a constant ($\epsilon$-independent) BL radius $r_0$.  (Note that we use the notation $x_0$ for the ISCO radius in the main body.)  The ``cracks'' in the throat illustrate infinite proper radial distance on a BL slice in the extremal limit.  They can also be interpreted as signaling the presence of three physically distinct extremal limits (App.~\ref{app:limits}).}
\label{fig:throats}
\end{figure}

We consider a non-extremal  ($\epsilon>0$) Kerr black hole and work with the dimensionless radial coordinate $x$ defined by
\begin{equation}\label{x}
x = \frac{r-r_+}{r_+},
\end{equation}
which places the event horizon at $x=0$.  The exterior of a non-extremal black hole has three important circular equatorial geodesics picked out by geometric considerations \cite{bardeen-press-teukolsky1972}: the ISCO (the marginally stable orbit), the innermost bound circular orbit (the marginally bound orbit) and the photon orbit or light ring.  As noted by \cite{bardeen-press-teukolsky1972}, the (BL or $x$) coordinate radii of these orbits approach that of the horizon as $\epsilon \rightarrow 0$.  The marginally bound and photon orbits go like $x\sim \epsilon$, while the ISCO approaches more slowly, being given to leading order in $\epsilon$ by
\begin{equation}\label{xISCO}
x_0 = 2^{1/3} \epsilon^{2/3}.
\end{equation}
Fig.~\ref{fig:throats} illustrates the properties of these orbits, and a formal discussion of their $\epsilon \rightarrow 0$ limits is given in App.~\ref{app:limits}.  While our focus is on the ISCO, our analysis holds for any orbit going like $x_0 \sim \epsilon^k$ with $0<k<1$.  Except where explicitly noted, all later formulae in this paper hold for such orbits.

Two other useful properties of a circular orbit are its angular velocity $\Omega$ and ``redshift factor" $g=e-\Omega l$ (where $e$ and $l$ are the particle's conserved energy and angular momentum per unit rest mass).  To leading order we have
\begin{align}\label{ISCOproperties}
\frac{\Omega - \Omega_H}{\Omega_H}  = -\frac{3}{4} x_0, 
\quad g  = \frac{\sqrt{3}}{4} x_0. 
\end{align}
The physical significance of $g$ is that a photon emitted by the particle with energy $\mathcal{E}$ is observed on the symmetry axis at infinity to have energy $g \mathcal{E}$.  This thought experiment illustrates how signals from the near-horizon region are redshifted away; in the case of the ISCO the observed energy vanishes as $x_0\sim \epsilon^{2/3}$ as extremality is reached.  This is the same scaling as the radiation from the particle orbit, our focus in this paper.

Note that as $\epsilon \rightarrow 0$ the horizon angular velocity and BL horizon radii go as
\begin{align}
\Omega_H & = 
\frac{1}{2M}\left( 1 - \epsilon \right). \label{OmegaH} \\ 
r_\pm & =M\left( 1 \pm \epsilon \right). 
\end{align}
Since $\Omega_H \rightarrow 1/(2M)$ like $\epsilon$, one may replace $\Omega_H$ with $1/(2M)$ in Eq.~\eqref{ISCOproperties}.

\section{Scalar Calculation}\label{sec:scalar}

We first define the problem at finite $\epsilon>0$.  We consider the scalar wave equation,
\begin{equation}\label{eq:scalar_field_eq}
g^{ab}\nabla_a \nabla_b \Phi = -4 \pi T,
\end{equation}
with source
\begin{align}\label{source}
T & = \frac{qg}{r_0^2}\delta(r-r_0) \delta (\theta - \pi/2) \delta(\phi - \Omega t).
\end{align}
Here $q$ is a constant called the scalar charge,  $g$ is the redshift factor \eqref{ISCOproperties}, $r_0$ is the BL coordinate radius of the ISCO, and $\Omega$ is its angular velocity. The source has a mode expansion,
\begin{equation}
T = \frac{q g}{r_0^2} \sum_{\ell, m} \delta(r-r_0)S_{\ell m}(\pi/2) S_{\ell m}(\theta)  e^{im(\phi - \Omega t)},
\end{equation}
where $\ell$ ranges from $0$ to $\infty$ and $m$ ranges from $-\ell$ to $\ell$.  Only these modes will be excited in the field, which we similarly decompose as
\begin{equation}\label{phidecomp}
\Phi = \sum_{\ell,m} \Phi_{\ell m} = \sum_{\ell,m} R_{\ell m}(r) S_{\ell m}(\theta) e^{im(\phi - \Omega t)}.
\end{equation}
The $S_{\ell m}$ satisfy the spheroidal harmonic differential equation,
\begin{equation}\label{Seqn}
\left[ \frac{\partial_\theta(\sin \theta \partial_\theta)}{\sin \theta}  + K_{\ell m} - \frac{m^2}{\sin^2 \theta} - a^2 m^2 \Omega^2 \sin^2 \theta \right] S_{\ell m} = 0.
\end{equation}
Solutions regular at the poles are labeled by $\ell$ and $m$ with an associated eigenvalue $K_{\ell m}$.  We normalize them so that $\int\sin \theta d\theta  S^2 =1$.

In terms of the dimensionless coordinate $x$ \eqref{x}, the radial functions satisfy
\begin{align}
 x (x+\sigma) R''(x)+&(2x+\sigma) R'(x) +V R(x) = \label{Teuk} \\ & \frac{- 2 q g}{r_+}S_{\ell m}({\pi}/{2}) \delta(x-x_0) , \nonumber
\end{align}
with
\begin{equation}
V = \frac{(r_+ m \Omega x(x+2) + n\sigma /2)^2}{x(x+\sigma)} + 2 a m^2 \Omega - K,
\end{equation}
where we have introduced
\begin{equation}\label{sigman}
\sigma = \frac{r_+ - r_-}{r_+},
\quad n = 4mM\frac{\Omega-\Omega_H}{\sigma}.
\end{equation}
We have dropped the mode labels $\ell$ and $m$.  Eq.~\eqref{Teuk} is also the spin-zero Teukolsky equation \cite{teukolsky1973} with angular frequency $\omega=m\Omega$.

For the non-radiative $m=0$ modes, Eq.~\eqref{Teuk} can be solved exactly for any value of spin \cite{Warburton:2010eq}.  We focus on the radiative case, and for the remainder of the paper we assume $m \neq 0$.  In this case, the solutions to \eqref{Teuk} have asymptotic behaviors given by
\begin{align}
R(x) & \rightarrow {C^\infty} e^{i m \Omega r_+ x} x^{-1+i m\Omega r_+ \sigma/\epsilon} & \nonumber \\ & + {D^\infty} e^{-i m \Omega r_+ x} x^{-1-i m\Omega r_+ \sigma/\epsilon} ,  & x\rightarrow \infty \label{Rinf} \\
R(x) & \rightarrow C^H x^{-i n r_+ \sigma/(4M\epsilon)} & \nonumber \\ & + D^H x^{i  n r_+ \sigma/(4M\epsilon)} , & \quad x \rightarrow 0, \label{Rhor}
\end{align}
where $C^\infty$, $D^\infty$, $C^H$, and $D^H$ are (complex) constants.  We impose no incoming radiation from the past horizon or past null infinity,
\begin{equation}\label{noinc}
D^\infty = D^H = 0.
\end{equation}
From the properties of the differential equation \eqref{Teuk}, it is clear that this uniquely fixes the solution.  The observables we are interested in are the power radiated to infinity and down the event horizon.  These are given for each mode by
\begin{align}
\dot{\mathcal{E}}_{\infty} & = \frac{1}{2} r_+^2 m^2 \Omega^2 |C^\infty|^2 \label{GiveMeThePowerInf}\\
\dot{\mathcal{E}}_{H} & = M r_+ m^2 \Omega(\Omega-\Omega_H) |C^H|^2.\label{GiveMeThePowerHor}
\end{align}
This defines the problem for every $\epsilon>0$.

The need for a matched expansion to study the $\epsilon \rightarrow 0$ limit can be seen at the level of the differential equation.  Naively setting $\epsilon=0$ in \eqref{Teuk} and solving, one finds that the solutions go as $x^{h-1}$ and $x^{-h}$ near $x = 0$, rather than the oscillatory behavior \eqref{Rhor} of the finite-$\epsilon$ equation.  Thus the $\epsilon=0$ equation cannot satisfy the boundary conditions of the problem, the hallmark of a singular perturbation problem.

\subsection{Near-extremal Simplification}
We first make some simplifications using $\epsilon \ll 1$.  The angular equation \eqref{Seqn} becomes
\begin{equation}\label{SeqnSimp}
\left[ \frac{\partial_\theta(\sin \theta \partial_\theta)}{\sin \theta}  + K - m^2 \left( \frac{1}{\sin^2 \theta}+ \frac{1}{4} \sin^2 \theta \right) \right] S = 0,
\end{equation}
which is independent of the frequency $\Omega$ and hence independent of $\epsilon$. To leading order Eq.~\eqref{sigman} becomes
\begin{align}
\sigma& =2\epsilon \\
n & = -\frac{3}{4} m x_0 \epsilon^{-1},  \label{n2} 
\end{align}
where we have used Eq.~\eqref{ISCOproperties} to get the second relation. For the ISCO we see that $n$ diverges as $n\sim\epsilon^{-1/3}$. In App.~\ref{app:limits} $n$ is related to the frequency conjugate to the time of the near-horizon metric. The radial equation \eqref{Teuk} becomes
\begin{align}\label{TeukSimp}
 & x (x+2\epsilon) R''+2(x+\epsilon) R'+ \hat{V} R = Nx_0 \delta(x-x_0) ,
\end{align}
where we introduce
\begin{equation}\label{alphahat}
N =  -\frac{\sqrt{3}}{2}\frac{ q }{M} S_{\ell m}({\pi}/{2})
\end{equation}
and
 \begin{equation}
\hat{V} = \frac{( \frac{1}{2}m x(x+2)- n \epsilon)^2}{x(x+2\epsilon)} +m^2  - K.
\end{equation}
We can also simplify \eqref{GiveMeThePowerInf}, \eqref{GiveMeThePowerHor} using \eqref{ISCOproperties},
\begin{align}
\dot{\mathcal{E}}_{\infty} & =  \frac{1}{8} m^2 |C^\infty|^2 \label{Pinf} \\
\dot{\mathcal{E}}_{H} & = -\frac{3}{16}  x_0 m^2 |C^H|^2. \label{Phor}
\end{align}
Notice that $\dot{\mathcal{E}}_{H}<0$, indicating that these modes are superradiant.

\subsection{Matched Asymptotic Expansions Overview}

For $x \gg x_0$, Eq.~\eqref{TeukSimp} becomes
\begin{equation}\label{fareqn}
 x ^2 R''+2x R' +[m^2 (2+x+ x^2/4) - K]R = 0.
\end{equation}
(Note that $x_0 \sim n \epsilon$ by \eqref{n2}.)  This is the ``far'' equation and its solutions will carry the label ``far''.  For $x \ll 1$ Eq.~\eqref{TeukSimp} instead becomes
\begin{align}\label{neareqn}
&  x (x+2\epsilon) R''+2(x+\epsilon) R' \\  & \quad + \left[\frac{( m x +  n \epsilon )^2}{x(x+2\epsilon)} +m^2-K \right] R = N x_0 \delta(x-x_0). \nonumber
 \end{align}
Eq.~\eqref{neareqn} is the ``near'' equation and its solutions will carry the label ``near''.  The equations agree when $x_0 \ll x \ll 1$, becoming
\begin{equation}\label{overlapeqn}
 x^2 R''+2x R' +[2m^2 - K]R = 0.
\end{equation}
This is the ``region of overlap'' and the solutions are
\begin{equation}\label{Roverlap}
R^{\rm overlap} = P x^{h-1} + Q x^{-h}
\end{equation}
for constants $P$ and $Q$, where $h$ is given in \eqref{h}.  This region corresponds to the $x\rightarrow 0$ behavior of solutions of the far equation \eqref{fareqn} and the $x \rightarrow \infty$ behavior of solutions to the near equation \eqref{neareqn}.  Thus each solution of \eqref{fareqn} or \eqref{neareqn} is characterized by values of $P$ and $Q$ obtained by looking at the appropriate asymptotic region.  A pair of solutions approximates a single smooth solution to Eq.~\eqref{TeukSimp} (and hence \eqref{Teuk}) when the solutions have the same $P$ and $Q$.

\subsection{Far Solutions}

The far equation \eqref{fareqn} is a confluent hypergeometric equation and its solutions can be written in a number of equivalent ways.  We parameterize the general solution by $P$ and $Q$,
\begin{align}
R^{\rm far} & = P x^{h-1} e^{- i m x /2} {}_1F_1\left(h+i m; 2h; i m x\right) \label{farsoln} \\
& \quad + Q x^{-h} e^{- i m x /2} {}_1F_1\left(1-h+i m; 2(1-h); i m x\right). \nonumber
\end{align}
That is, at small $x$ we have
\begin{equation}\label{farsmallx}
R^{\rm far} \rightarrow P x^{h-1} + Q x^{-h}, \quad x \rightarrow 0.
\end{equation}
Notice that the two solutions are related by $h\rightarrow1-h$.  For large $x$ the asymptotic behavior is
\begin{align}
R^{\rm far} & \rightarrow C^\infty e^{i m x/2} x^{-1+im}  \nonumber \\ & \quad + D^\infty e^{-i m x/2} x^{-1-im}, \qquad x\rightarrow \infty.\label{farlargex}
\end{align}
with
\begin{align}
C^\infty & = P \frac{(im)^{-h+im} \Gamma(2h)}{\Gamma(h+im)} + Q\frac{(im)^{h-1+im} \Gamma(2(1-h))}{\Gamma(1-h+im)} \nonumber \\
D^\infty & = P \frac{(-im)^{-h-im} \Gamma(2h)}{\Gamma(h-im)} + Q\frac{(-im)^{h-1-im} \Gamma(2(1-h))}{\Gamma(1-h-im)} \nonumber
\end{align}
To be outgoing at infinity we must have $D^\infty=0$ or
\begin{equation}\label{PQ}
P/Q = (-im)^{2h-1} \frac{ \Gamma(1-2h)\Gamma(h-im)}{\Gamma(2h-1) \Gamma(1-h-im)}. 
\end{equation}
In this case the coefficient $C^\infty$ is given by
\begin{align}
C^\infty& =Q\,\frac{\Gamma(2-2h)}{\Gamma(1-h+im)}(im)^{h-1+im} \nonumber \\ & \quad \times
\left[1- \frac{(-im)^{2h-1}}{(im)^{2h-1}}\frac{\sin[\pi(h+im)]}{\sin[\pi(h-im)]}\right]  \label{Cinfpureout} \\
& =  -Q\,(-1)^{- \textrm{sign}(m)\, h} \frac{\Gamma(h-im)}{\Gamma(2h-1)} e^{\pi|m|} (im)^{h-1+im}. \nonumber
\end{align}

\subsection{Near Solutions}
The near equation \eqref{neareqn} is a hypergeometric equation.  We will work with the following two linearly independent homogeneous solutions,\footnote{This choice expedites writing the answer for the horizon flux in a form that makes manifest the scaling for small $\epsilon$.}
\begin{align}
R^{\rm near}_{\rm in} & = x^{-\frac{i n}{2}} \left(\frac{x}{2\epsilon}+1\right)^{i \left(\frac{n}{2}-m\right)} \nonumber \\ & \quad \times
  {}_2F_1\left(h-i m,1-h-i m;1-i n;-\frac{x}{2\epsilon}\right) \label{Rnearin} \\
R^{\rm near}_{\rm N}  & = x^{-h}  \left(\frac{2\epsilon}{x}+1\right)^{i \left(\frac{n}{2}-m\right)}\nonumber \\ & \quad \times
  {}_2F_1\left(h-i m,h+i(n-m);2h;-\frac{2\epsilon}{x}\right) \label{RnearN}
\end{align}
The asymptotic behaviors are
\begin{align}
R_{\rm in}^{\rm near} &\to x^{-in/2} \quad ~& \textrm{for $x\to 0$}\,,\label{Rnearin0} \\
 &\to A\, x^{h-1}+B\, x^{-h} \quad~& \textrm{for $x\to \infty$}\,,\\
R_{\rm N}^{\rm near} &\to C\, x^{-in/2}+D\, x^{in/2}~& \textrm{for $x\to 0$}\,,\\
 &\to  x^{-h} \quad~& \textrm{for $x\to \infty$}\,,\label{say}
\end{align}
where
\begin{align}
A & =\frac{\Gamma(2h-1)\Gamma(1-in)}{\Gamma(h-im)\Gamma(h-i(n-m))}(2\epsilon)^{1-h-\frac{in}{2}}\,, \label{A} \\
B & =\frac{\Gamma(1-2h)\Gamma(1-in)}{\Gamma(1-h-im)\Gamma(1-h-i(n-m))}(2\epsilon)^{h-\frac{in}{2}}\,, \label{B} \\
C &= \frac{\Gamma (2 h) \Gamma (i n) }{\Gamma (h+i m) \Gamma (h+i(n-m) )}(2\epsilon)^{-h+\frac{in}{2}}\,, \label{C} \\
D&=\frac{\Gamma (2 h) \Gamma (-i n) }{\Gamma (h-i m) \Gamma (h-i(n-m))}(2 \epsilon )^{-h-\frac{in}{2}}\,. \label{D}
\end{align}
The ``in'' solution is purely ingoing at the horizon, while the ``N'' solution has only the $x^{-h}$ falloff at large $x$.  Here ``N'' stands for Neumann, which is the terminology used in \cite{porfyriadis-strominger2014}.\footnote{The reason for this terminology is that for real $h$ the falloff $x^{-h}$ is subdominant compared to $x^{1-h}$.}  The Wronskian $W$ is given by
\begin{equation}\label{Wronskian}
x(x+2\epsilon) W(R_{\rm in}^{\rm near},R_{\rm N}^{\rm near}) = (1-2h)A.
\end{equation}
From the properties of the differential equation, the combination on the LHS above is known to be independent of $x$ and may therefore be easily computed at large $x$.

\subsection{Up solution}

We now consider the solution with pure outgoing radiation at infinity, conventionally called the ``up'' solution.  The normalization is arbitrary and we will choose
\begin{equation}\label{Rnearup}
R^{\rm near}_{\rm up} = R^{\rm near}_{\rm in} + \alpha R^{\rm near}_{\rm N}
\end{equation}
in the near zone.  At large $x$ we have $R^{\rm near}_{\rm up}=A x^{h-1}+(B+\alpha)x^{-h}$.  Matching to \eqref{farsmallx}, we have $P=A$ and $Q=B+\alpha$.  We can thus write
\begin{align}
\alpha & = B\left(\frac{A}{B} \frac{Q}{P}-1\right) = B(1/b-1), \label{alpha}
\end{align}
where $B$ is given in \eqref{B}, and from \eqref{PQ}, \eqref{A}, and \eqref{B} we can compute $b\equiv(B/A)(P/Q)$ to be
\begin{align}\label{b}
 b & = (-i m)^{2h-1} \frac{\Gamma(1-2h)^2\Gamma(h-i m)^2}{\Gamma(2h-1)^2 \Gamma(1-h-i m)^2} \nonumber \\ & \qquad  \times \frac{\Gamma(h+i(m-n))}{\Gamma(1-h+i(m-n))} (2\epsilon)^{2h-1}.
\end{align}
The up solution in the far zone is given by Eq.~\eqref{farsoln} with
\begin{align}
P = A , \quad \quad
Q = B+\alpha =B/b.\label{Qup}
\end{align}
The behavior near infinity (which controls the outgoing radiation) is given by plugging $Q=B/b$ into \eqref{Cinfpureout}.  We thus have
\begin{align}\label{Cinfup}
C^{\infty}_{\rm up} = -\frac{B}{b} (-1)^{-\textrm{sign}(m)\,h} \frac{\Gamma(h-im)}{\Gamma(2h-1)} e^{\pi |m|} (im)^{h-1+im}.
\end{align}

\subsection{Retarded Solution}

To construct the retarded solution we demand pure ingoing at the horizon, pure outgoing at infinity, and the proper match at the delta-function source at $x=x_0$ in Eq.~\eqref{neareqn}.  This is given by
\begin{align}
R^{\rm near}_{\rm ret}(x)  & = N x_0\frac{R^{\rm near}_{\rm in}(x_<)R^{\rm near}_{\rm up}(x_>)}{x(x+2\epsilon)W[R^{\rm near}_{\rm in}(x),R^{\rm near}_{\rm up}(x)]} \nonumber \\
& = \frac{N x_0}{\alpha A(1-2h)}R^{\rm near}_{\rm in}(x_<)R^{\rm near}_{\rm up}(x_>),
\end{align}
where $x_<$ and $x_>$ are the lesser and greater of $x$ and $x_0$, respectively.  We have used \eqref{Rnearup} and \eqref{Wronskian} to evaluate the Wronskian.

\subsection{Large-$n$ Asymptotics}

Thus far we have considered exact solutions of the near equation \eqref{neareqn}, where $\epsilon$ (and hence $x_0$ and $n$) is treated as finite.
We now simplify further using the smallness of $\epsilon$, which by \eqref{n2} corresponds to large $n$.

We first simplify the expressions for $A$, $B$, $\alpha$, and $b$.  For this we need the following asymptotic approximation,
\begin{equation}
\frac{\Gamma(p+z)}{\Gamma(q+z)} = z^{p-q} , \quad z \rightarrow \infty,\label{gammaasymptotics}
\end{equation}
which holds for any complex $p,q,z$.  From \eqref{A} and \eqref{B} we have
\begin{align}
A &= \frac{\Gamma(2h-1)}{\Gamma(h-im)}(-in)^{-im}(2\epsilon)^{-in/2}\left(\tfrac{3}{2}imx_0\right)^{1-h} \label{Alargen} \\
B & =\frac{\Gamma(1-2h)}{\Gamma(1-h-im)}(-in)^{-im}(2\epsilon)^{-in/2}\left(\tfrac{3}{2}imx_0\right)^{h}, \label{Blargen}
\end{align}
and from \eqref{alpha} and \eqref{b} we have
\begin{align}
\alpha & = B(1/b-1) \label{alphalargen}\\
b &= \frac{\Gamma(1-2h)^2}{\Gamma(2h-1)^2}\frac{\Gamma(h-i m)^2}{\Gamma(1-h-i m)^2} (3m^2x_0/2)^{2h-1}.\label{blargen}
\end{align}
(We repeat Eq.~\eqref{alpha} for convenience.) In obtaining these equations we have used that $n \epsilon = -3mx_0/4$ from Eq.~\eqref{n2}.

When the radial functions are evaluated at $x=x_0$, the ${}_2F_1$ hypergeometrics in Eqs.~\eqref{Rnearin} and \eqref{RnearN} reduce to Whittaker $W$ and $M$ functions via the confluence identities:
\begin{align}
W_{\nu,\mu}(z)  & = \lim_{c \rightarrow \infty} {}_2F_1(\mu - \nu+\tfrac{1}{2},-\mu - \nu+\tfrac{1}{2};c;1-\tfrac{c}{z}) \nonumber \\ & \quad \quad \times e^{-z/2} z^{\nu} \label{Wdef} \\
M_{\nu,\mu}(z)  & = \lim_{b \rightarrow \infty} {}_2F_1(\mu - \nu+\tfrac{1}{2},b;1+2\mu;\frac{z}{b}) \nonumber \\ & \quad \quad \times e^{-z/2} z^{\mu+\frac{1}{2}}. \label{Mdef}
\end{align}
Specifically, we have
\begin{align}
 R^{\rm near}_{\rm in}(x_0) & = x_0^{-\frac{i n}{2}} \left(\frac{x_0}{2\epsilon}+1\right)^{i \left(\frac{n}{2}-m\right)} \nonumber \\ & \quad \times e^{3i m/4} (3 i m/2)^{-im} W_{im,h-\frac{1}{2}}(3 i m /2) \label{Rnearinlargen} \\
   R^{\rm near}_{\rm N}(x_0) & = (3 i m x_0 /2)^{-h} \left(\frac{2\epsilon}{x_0}+1\right)^{i \left(\frac{n}{2}-m\right)} \nonumber \\ & \quad \times e^{3i m/4} M_{im,h-\frac{1}{2}}(3 i m /2),\label{RnearNlargen}
\end{align}
where we have again used that $n \epsilon = -3mx_0/4$.

\subsection{Horizon flux}
To compute the power radiated down the event horizon we need to examine the $x \rightarrow 0$ behavior of our solution and extract the coefficient $C^H$ defined in \eqref{Rhor}.  For $x<x_0$ the solution is given by
\begin{align}
R^{\rm near}_{\rm ret}(x) & = \frac{N x_0}{\alpha A (1-2h)}R^{\rm near}_{\rm in}(x)R^{\rm near}_{\rm up}(x_0).
\end{align}
Thus the horizon coefficient is
\begin{align}
C^{\rm H}_{\rm ret} & = \frac{N x_0}{\alpha A(1-2h)}C^{\rm H}_{\rm in}R^{\rm near}_{\rm up}(x_0),
\end{align}
where $C^{\rm H}_{\rm in}$ is the horizon coefficient for the in solution.  However, from \eqref{Rnearin0} we have simply $C^{\rm H}_{\rm in}=1$.  Using Eqs.~\eqref{Rnearup} and \eqref{alphalargen}, a more convenient expression is
\begin{align}
C^{\rm H}_{\rm ret} & = \frac{N x_0}{AB (1-2h)}\left(B R^{\rm near}_{\rm N}(x_0) + \frac{b}{1-b}R^{\rm near}_{\rm in}(x_0)\right).
\end{align}
Using Eqs.~\eqref{alphahat}, \eqref{Alargen}, \eqref{Blargen}, \eqref{blargen}, \eqref{Rnearinlargen}, and \eqref{RnearNlargen} and simplifying, we find
\begin{align}
C^{\rm H}_{\rm ret} & = N\frac{2i}{3m}e^{3 i m/4}(-in)^{im}(2 \epsilon)^{i n/2}  \left(1+\frac{2\epsilon}{x_0}\right)^{i \left(\frac{n}{2}-m\right)}  \nonumber \\ & \times \frac{\Gamma(h-im)}{\Gamma(2h)} \left( \mathscr{M} + \frac{b}{1-b}\frac{\Gamma(1-h-im)}{\Gamma(1-2h)} \, \mathscr{W} \right), \label{CHret}
\end{align}
where we introduce
\begin{align}\label{MW}
\mathscr{M} = M_{im,h-\frac{1}{2}}\!\left(\frac{3 i m}{2}\right), \  \ \
\mathscr{W} = W_{im,h-\frac{1}{2}}\!\left(\frac{3 i m}{2}\right).
\end{align}
Squaring and plugging into \eqref{Phor} gives the power radiated,
\begin{align}
\dot{\mathcal{E}}_{H} & = - \frac{q^2}{16M^2} \, x_0 \, e^{-\pi |m|} S\!\left(\tfrac{\pi}{2}\right)^2 \left|\frac{\Gamma(h-im)}{\Gamma(2h)}\right|^2 \nonumber \\ & \quad \times \left| \mathscr{M} + \frac{b}{1-b}\frac{\Gamma(1-h-im)}{\Gamma(1-2h)} \, \mathscr{W} \right|^2. \label{GloriousHorizonScalar}
\end{align}
The energy flux down the horizon scales as $x_0$, so that for the ISCO it scales as $\epsilon^{2/3}$.  The formula for $b$ was given in Eq.~\eqref{blargen}.  When $h$ has an imaginary part, $b$ is order unity and oscillatory in $\epsilon$, causing $\dot{\mathcal{E}}_H$ to have small oscillations.  When $h$ is real, $b \ll 1$ and the entire term proportional to $\mathscr{W}$ drops out at the leading order, making there be no oscillations.

\subsection{Infinity flux}
For $x>x_0$ the solution is given in the near-zone by
\begin{align}
R^{\rm near}_{\rm ret}(x) & = \frac{N x_0}{\alpha A(1-2h)}R^{\rm near}_{\rm in}(x_0)R^{\rm near}_{\rm up}(x).
\end{align}
This solution is valid in the near-zone, with $x\rightarrow \infty$ corresponding to the overlap region rather than asymptotic infinity.  To determine the behavior near asymptotic infinity one has to match to solutions of the far region.  However, this has already been done when constructing the ``up'' solution.  Thus the retarded solution near infinity is determined by
\begin{align}
C^{\infty}_{\rm ret} & = \frac{N x_0}{\alpha A(1-2h)}R^{\rm near}_{\rm in}(x_0)C^\infty_{\rm up}.
\end{align}
Using Eq.~\eqref{Cinfup} for $C^\infty_{\rm up}$ gives
\begin{align}
C^{\infty}_{\rm ret} & = N x_0\frac{1}{1-b}\frac{R^{\rm near}_{\rm in}(x_0)}{A} \nonumber \\ & \times   (-1)^{-\textrm{sign}(m)\,h} \frac{\Gamma(h-im)}{\Gamma(2h)} e^{\pi|m|} (im)^{h-1+im}.
\end{align}
Using Eqs.~\eqref{Rnearinlargen} and \eqref{Alargen} and simplifying gives
\begin{align}
C^{\infty}_{\rm ret} & = N x_0^h\, e^{3 i m/4} \left(\frac{2\epsilon}{x_0}+1\right)^{i \left(\frac{n}{2}-m\right)} \frac{2h-1}{1-b} \frac{\Gamma(h-im)^2}{\Gamma(2h)^2}\nonumber \\ & \times   (-1)^{-\textrm{sign}(m)\,h}  e^{\pi|m|} (im)^{h-1+im}(3 i m /2)^{h-1} \, \mathscr{W},\label{Cinfret}
\end{align}
where $\mathscr{W}$ was defined in \eqref{MW}.  Squaring and plugging in to \eqref{Pinf} gives the infinity flux,
\begin{align}
\dot{\mathcal{E}}_{\infty} & = \frac{q^2}{24M^2} (3 m^2 x_0 /2)^{2 \textrm{Re}[h]} m^{-2} e^{\pi|m| } S\!\left(\tfrac{\pi}{2}\right)^2  \nonumber  \\ & \qquad \times \left| \frac{2h-1}{1-b} \frac{\Gamma(h-im)^2}{\Gamma(2h)^2} \mathscr{W} \right|^2. \label{GloriousInfinityScalar}
\end{align}
Recall that $b$ is given in Eq.~\eqref{blargen} and $\mathscr{W}$ is given in Eq.~\eqref{MW}.  The energy flux to infinity scales as $x_0^{2 \textrm{Re}[h]}$, so that for the ISCO it scales as $\epsilon^{(4/3) \textrm{Re}[h]}$.  The dominant modes are when $h$ has an imaginary part, in which case $\textrm{Re}[h]=1/2$ and $b \sim 1$ with oscillations.  The modes with real $h$ are subdominant with $b \ll 1$ and no oscillations at leading order.

\subsection{Agreement with extremal calculation}\label{sec:agreement}

Ref. \cite{porfyriadis-strominger2014} solved the physically distinct problem of the radiation from a particle orbiting at a radius $x_0 \ll 1$ in precisely extremal Kerr.  Remarkably, our final answer \eqref{GloriousInfinityScalar} for the flux to infinity agrees \textit{precisely} with the analogous answer (3.54) of \cite{porfyriadis-strominger2014} when we identify the $x$ coordinates \eqref{x} in extremal and near-extremal Kerr.\footnote{Ref.~\cite{porfyriadis-strominger2014} uses the notation $r$ for our $x$ and $\lambda$ for our $q$.}  In \cite{porfyriadis-strominger2014} the horizon flux was not explicitly given for the retarded solution, but performing the calculation shows perfect agreement as well. Near infinity we can also sensibly compare the detailed radiation pattern, which agrees as well: The asymptotic behavior of the retarded field is given by \eqref{Cinfret} together with \eqref{farlargex} and $D^{\infty}=0$, which differs from the expressions in \cite{porfyriadis-strominger2014} only by the phase $\exp[3 i m /4] (2\epsilon/x_0+1)^{i(n/2-m)}$.  This phase is of the form $\exp[i m f(\epsilon)]$ and hence can be eliminated at any fixed $\epsilon$ by the redefinition $\phi \rightarrow \phi - f(\epsilon)$.

For real $h$ the horizon flux \eqref{GloriousHorizonScalar} is in perfect agreement with the particle number flux (3.27) of \cite{porfyriadis-strominger2014}. Using the Kerr/CFT dictionary, this flux was calculated independently in the CFT as the appropriate transition rate induced on the state of the system due to coupling to a source dual to the orbiting particle (Eq.~(3.40) of \cite{porfyriadis-strominger2014}). Note, however, that the boundary conditions used for deriving (3.27) of \cite{porfyriadis-strominger2014} assumed Neumann falloff of the near solution (rather than the ``up'' falloff of \eqref{Rnearup}). The boundary conditions used here for the retarded solution were termed ``leaky boundary conditions'' in \cite{porfyriadis-strominger2014} because they allow radiation to leak out of the near region and reach future null infinity. However, as we saw in the previous section, for real $h$, the radiation that leaks to infinity here is subdominant and the CFT calculations of \cite{porfyriadis-strominger2014} still account for the flux down the horizon at leading order.

\section{Gravitational Case}\label{sec:gravity}
The problem of the gravitational radiation from a particle of mass $m_0$ orbiting on the near-extremal ISCO can be solved in a manner precisely analogous to the scalar calculation of Sec.~\ref{sec:scalar}.  However, given the agreement in the scalar case with the analogous calculation of \cite{porfyriadis-strominger2014} (Sec.~\ref{sec:agreement}), we can instead obtain analytic expressions by postulating agreement in the gravitational case as well.   The expression for the gravitational flux at infinity is given in Eq.~(4.41) of \cite{porfyriadis-strominger2014}. The flux at the horizon was not computed in \cite{porfyriadis-strominger2014}, but it is a straightforward exercise to do so.  Identifying these expressions using the $x$-coordinate radius of the particle produces expressions for the power radiated in our near-extremal ISCO problem. We confirm these expressions numerically.

We now present these results.  In some formulae we use the variable $s=-2$ (the spin) to emphasize similarity to the scalar case $s=0$.  For each mode $\ell\geq 2$ and $|m|\leq\ell$ the fluxes are given by
\begin{widetext}
\begin{align}
\dot{\mathcal{E}}_{H}&=-{m_0^2\over 2^6 3^2 M^2} {1\over |\mathcal{C}|^2} \, x_0 \, e^{-\pi|m| } {|\Gamma(h-im-s)|^2\over |\Gamma(2h)|^2} \left|\mathscr{M}_s+\frac{b_s} {1-b_s}\frac{\Gamma(1-h-im-s)}{\Gamma(1-2h)}\mathscr{W}_s \right|^2 \, \label{GloriousMagnificentHorizonGravity}\\
\dot{\mathcal{E}}_{\infty} &= {m_0^2\over 2^5 3^3 M^2} (3m^2x_0/2)^{2\mathrm{Re}[h]} m^{-2} e^{\pi|m|} \left| \frac{2h-1}{1-b_s} \frac{\Gamma(h-im-s)\Gamma(h-im+s)}{\Gamma(2h)^2}\mathscr{W}_s\right|^2, \label{GloriousMagnificentInfinityGravity}
\end{align}
where
\begin{align}
b_s & = \frac{\Gamma(1-2h)^2}{\Gamma(2h-1)^2}\frac{\Gamma(h-i m-s)}{\Gamma(1-h-i m-s)}  \frac{\Gamma(h-i m+s)}{\Gamma(1-h-i m+s)} \left(\frac{3m^2x_0}{2}\right)^{2h-1} \label{bhat} \\
\mathscr{M}_s &=2\left[(h^2-h+6-im)S+4(2i+m)S'-4S''\right]M_{im-2,h-\half}(3im/2)\notag\\
&\qquad-(h-2+im)\left[(4+3im)S-8iS'\right]M_{im-1,h-\half}(3im/2)\,\label{Mhat}\\
\mathscr{W}_s &=2\left[(h^2-h+6-im)S+4(2i+m)S'-4S''\right]W_{im-2,h-\half}(3im/2)\notag\\
&\qquad+\left[(4+3im)S-8iS'\right]W_{im-1,h-\half}(3im/2)\,. \label{What} \\
|{\mathcal C}|^2& = [(-2 + h)^2 + m^2][(-1 + h)^2 + m^2][h^2 + m^2][(1 + h)^2 + m^2]. \label{C} 
\end{align}
The spin-2 spheroidal harmonics $S(\theta)$ and their eigenvalues $K$ are defined to be the regular solutions to
\begin{equation}
\left[\frac{\partial_\theta(\sin\theta\,\partial_\theta)}{\sin\theta} +K^s_{\ell m}-\frac{m^2+s^2+2ms\cos\theta}{\sin^2\theta}- \frac{m^2}{4}\sin^2\theta-ms\cos\theta\right]S^s_{\ell m}=0,
\end{equation}
normalized so that $\int \sin\theta d\theta S^2=1$. In Eqs.~\eqref{Mhat} and \eqref{What}, a prime represents a  $\theta$-derivative and the harmonics are evaluated at the equator $\theta=\pi/2$.  The spin-2 spheroidal harmonics and their eigenvalues are not native in \textit{Mathematica}, but are straightforward to compute using (e.g.) the spectral method of \cite{Hughes:1999bq}.  We provide a notebook online \cite{Warburton:website} that implements this method and evaluates the complete analytic flux formulae.
\end{widetext}
Eq.~\eqref{bhat} for $b_s$ generalizes the scalar expression \eqref{b} for $b$, reducing to that expression when $s=0$.  Eqs.~\eqref{Mhat} and ~\eqref{What} for $\mathscr{M}_s$ and $\mathscr{W}_s$ generalize $\mathscr{M}$ and $\mathscr{W}$ in the sense that they play analogous roles in the expressions for the energy flux, but do not reduce to their scalar counterparts in any direct sense.  The formula for $|\mathcal{C}|^2$ is the same as Eq.~(3.23) of \cite{teukolsky-press1974} (specialized to the modes appearing in our calculation); it accounts for relating spin $\pm 2$ quantities in the Newman-Penrose formalism.

As in the scalar case,  $b_s \ll 1$ for real $h$, while $b_s \sim 1$ (and is oscillatory) for complex $h$. The character of the fluxes is thus precisely analogous to the scalar case; we refer the reader to the text below \eqref{GloriousHorizonScalar} and \eqref{GloriousInfinityScalar} for discussion.

\section{Numerical Results}\label{sec:numerical}

In this section we describe two new numerical codes and the comparison of their results with our analytic flux formulae. It was necessary to construct these new codes as previously developed software did not work sufficiently close to extremality to allow for a clear comparison with the analytic results. Typically the maximum spin achievable in these older codes was around $\epsilon\simeq10^{-3}$, equivalently $a\simeq0.9999995M$ \cite{Warburton:2010eq,Hughes:1999bq}. The two new codes we present here mark significant improvements over previous technology, allowing us to work all the way down to $\epsilon=10^{-13}$ or $a=0.999999999999999999999999995M$.

Our new codes feature three key improvements. First, the codes are implemented in \textit{Mathematica}, which allows us to work beyond standard machine precision. Second, for the gravitational case we derive new, more accurate, asymptotic approximations used for boundary conditions. Third, again in the gravitational case, we employ a simpler matching procedure to construct inhomogeneous solutions to the Teukolsky equation than that presented in Ref.~\cite{Hughes:1999bq}.  We briefly discuss these points in the following subsection before presenting the comparison between the numerical and analytic results in Sec.~\ref{sec:results_comparison}.

\subsection{Numerical Implementation}

\begin{figure*}
\centering
\subfigure[$\,\ell=2,m=2$]{\label{fig:22}\includegraphics[width=0.49\textwidth]{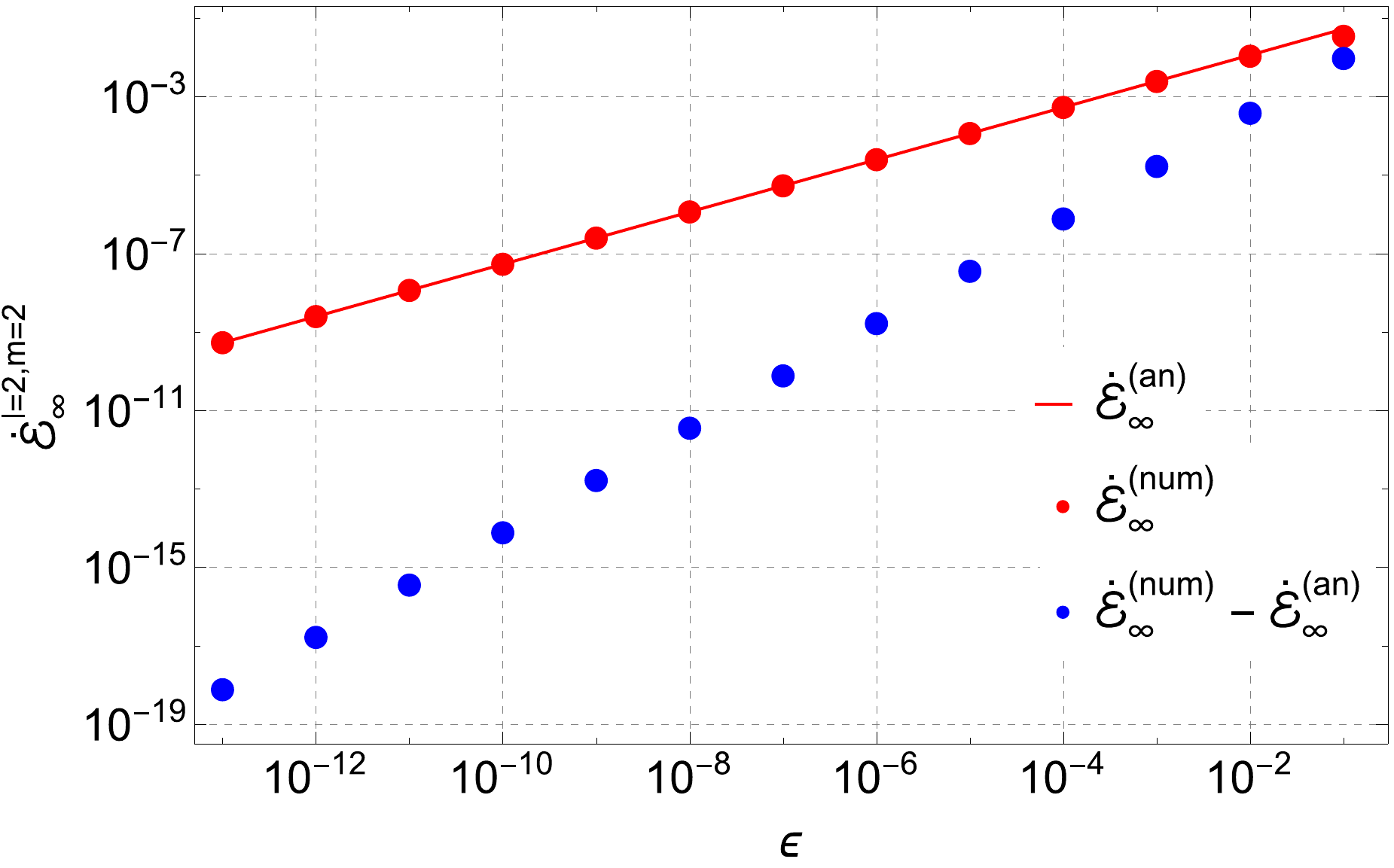}}\hspace{2mm}
\subfigure[$\,\ell=2,m=1$]{\label{fig:21}\includegraphics[width=0.49\textwidth]{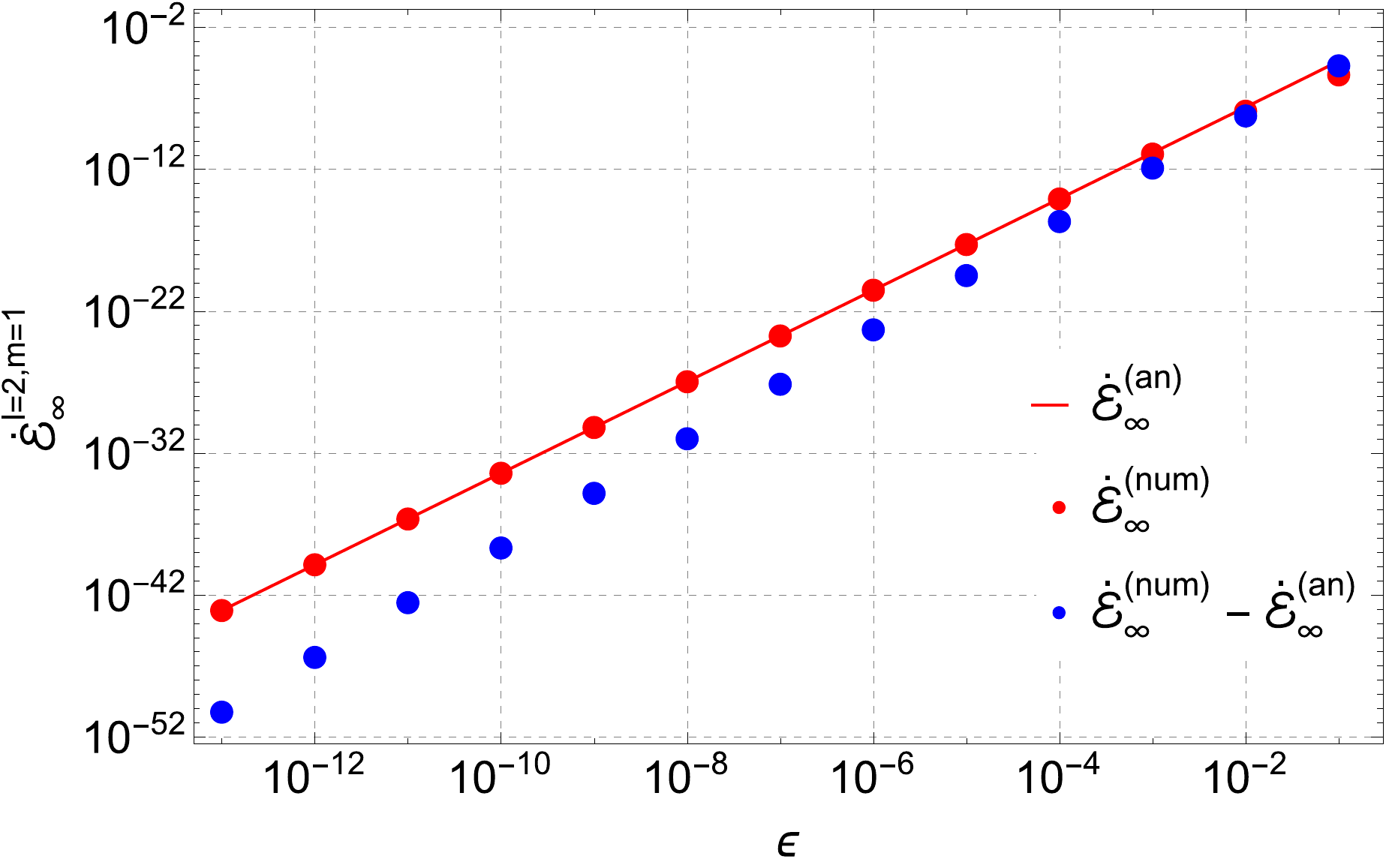}}
\caption{Energy flux to infinity for the 2-2 and 2-1 modes in the gravitational case. The analytic results and numerical results are denoted by $\dot{\mathcal{E}}^\text{(an)}_\infty$ and $\dot{\mathcal{E}}^\text{(num)}_\infty$, respectively. The 2-2 mode has $h\simeq1/2 + 2.050928i$ and hence an exponent of $p=2/3$, while the 2-1 mode has $h\simeq2.419070$ and an exponent of $p\simeq3.225427$.  The 2-2 mode also has oscillations, too small to be seen on this scale but clearly visible in Fig.~\ref{fig:oscillations}. Note that the results in this figure have been adimensionalized so that $\dot{\mathcal{E}}^\text{(here)}\equiv (M/m_0)^2\dot{\mathcal{E}}$.}
\label{fig:infFlux}
\end{figure*}

\begin{figure}
\includegraphics[width=0.48\textwidth]{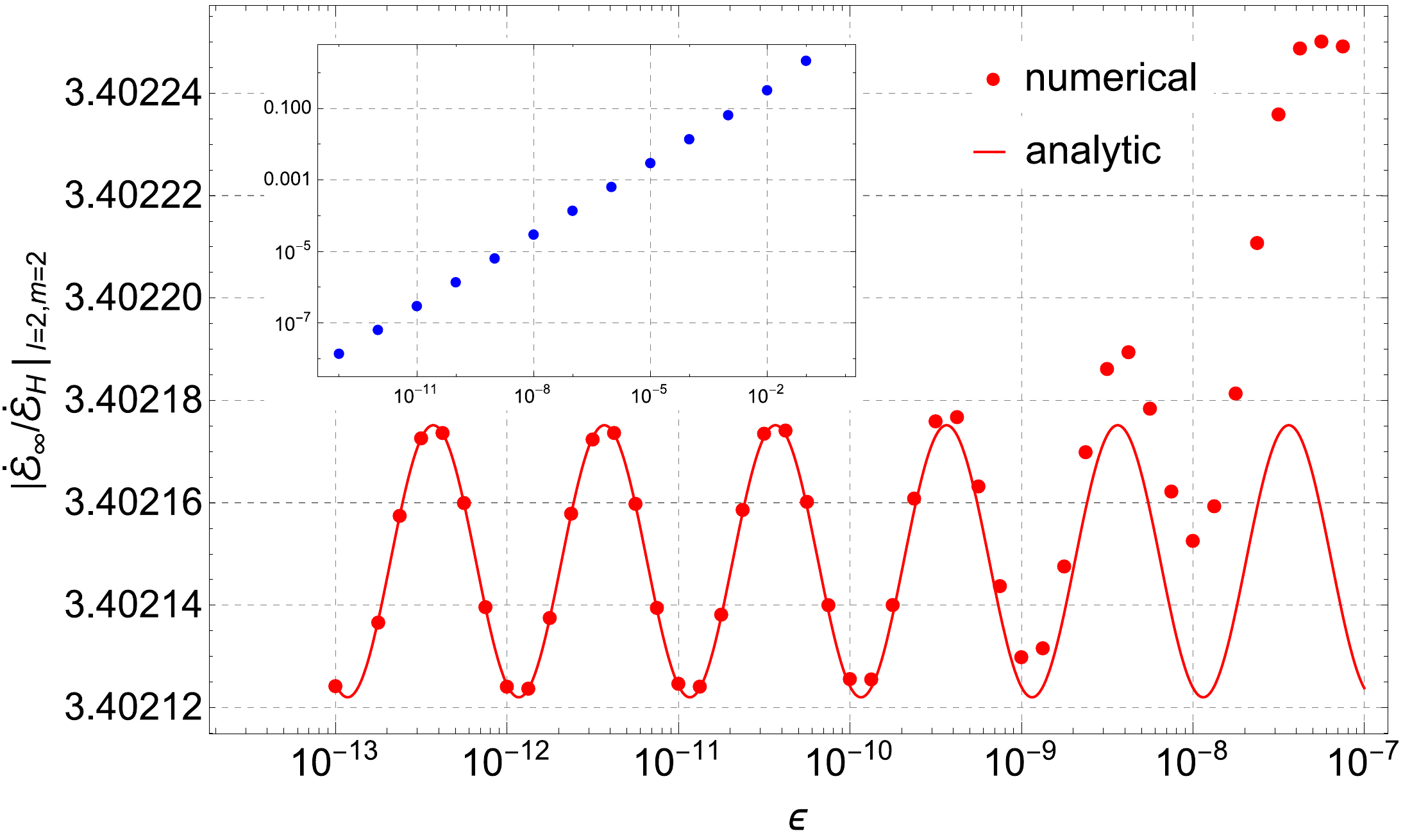}
\caption{The ratio of the infinity flux to horizon flux for the 2-2 mode in the gravitational case. The oscillations occur in both the infinity and horizon fluxes, but are dominated by the $\epsilon^{2/3}$ scaling. The ratio removes the dominant scaling and makes the oscillations clear. The inset shows the absolute difference between the analytic and numerical results for the ratio of the fluxes and demonstrates that our numerical results are in good agreement with the analytic formula through to $\epsilon=10^{-13}$.}
\label{fig:oscillations}
\end{figure}

The scalar problem was defined at the start of Sec.~\ref{sec:scalar}.  The main task is to solve the spin-zero Teukolsky equation \eqref{Teuk} with boundary conditions \eqref{noinc} corresponding to no incoming radiation.  We use a reimplementation in Mathematica of the algorithm presented in Ref.~\cite{Warburton:2010eq}. Briefly, the steps are as follows: (i) construct unit normalized boundary conditions for the homogeneous solutions far from the particle; (ii) using these boundary conditions, numerically integrate the field equation to get $R_{\ell m}$ and its derivative at the particle; (iii) compute weighting coefficients for the inhomogeneous solutions via the variations of parameters method. As our source contains a delta-function this reduces to a matching procedure at the particle's radius. We refer the reader to Ref.~\cite{Warburton:2010eq} for the explicit details of each of these steps.\footnote{We correct a mistake in the horizon-side boundary conditions in Ref.~\cite{Warburton:2010eq}. The arXiv version has been updated to give the correct recursion relation.}

For gravitational perturbations we opt to solve the spin-two Teukolsky equation. In this scenario we cannot proceed exactly as in the scalar case as the `long-ranged' potential of the Teukolsky equation makes it numerically challenging to avoid contamination from modes with incoming radiation.  This is a well known problem which is neatly circumvented by transforming to a new variable as was first described by Sasaki and Nakamura \cite{Sasaki:1981sx}. Using the new variable the field equation has a `short-ranged' potential and, as such, is better suited to numerical treatment. Once the Sasaki-Nakamura radial function and its derivatives are computed at the particle, we transform back to the Teukolsky radial function and continue, as in the scalar case, to construct the inhomogeneous solutions via matching. From the matching coefficients we then extract the radiated fluxes. A similar procedure was carried out by Hughes \cite{Hughes:1999bq}.  In addition to implementing our code in Mathematica, we make some important improvements as we outline now.

The Sasaki-Nakamura equation takes the form \cite{Sasaki:1981sx}
\begin{align}\label{eq:SN}
	\frac{dX}{dr_*^2} -F(r)\frac{dX}{dr_*} - U(r) X(r) = 0,
\end{align}
where $r_*$ is the tortoise coordinate defined by $dr_*/dr = (r^2+a^2)/\Delta$ with $\Delta = r^2 - 2Mr +a^2$. The functions $F(r)$ and $U(r)$ are rather unwieldy and can be found in the Appendix B of Ref.~\cite{Hughes:1999bq}. Asymptotically, as the horizon and spatial infinity are approached, the `outer' and `inner' radial solutions behave as
\begin{align}
	X^\infty(r_*\rightarrow\infty) 	&\sim e^{i\omega r_*},		\\
	X^H(r_*\rightarrow-\infty) 		&\sim e^{-i(\omega - m \Omega_H) r_*},	\label{eq:horiz_asymp_form}
\end{align}
respectively. In our numerical procedure we must work on a finite radial domain. Let us denote the boundaries of this domain by $r_\text{in}$ and $r_\text{out}$. In order to construct suitable boundary conditions at $r_\text{in/out}$ we expand the above asymptotic forms with the following ansatz
\begin{align}
	X^\infty 	&= e^{i\omega r_*}\sum_{k=0}^{k^\infty_\text{max}} a^\infty_k (\omega r_\text{out})^{-k},		\label{eq:inf_expansion}\\
	X^H			&= e^{-i(\omega - m \Omega_H) r_*}\sum_{k=0}^{k^H_\text{max}} a^H_k (r_\text{in} - r_+)^k .
\end{align}
The $a_k^{\infty/H}$ coefficients of these series expansions are determined by substituting the expansions into the Sasaki-Nakamura equation \eqref{eq:SN} and solving for the resulting recursion relations. Explicitly finding the recursion relations is one point where our algorithm differs from previous work. For example, the first three coefficients in Eq.~\eqref{eq:inf_expansion} were explicitly computed in Ref.~\cite{Hughes:1999bq}; the first four coefficients were given Ref.~\cite{Shah:2012gu}. Both the aforementioned works just set $a_0^H=1$ and used only the first term in the horizon expansion, i.e., $k_{\text{max}}^H=0$. We find that using just these few terms in the boundary condition expansions is insufficient for computations around rapidly rotating black holes. With our recursion relations we can compute arbitrary numbers of coefficients which allows us to place the numerical boundaries for the inner and outer solutions further from the horizon and closer to the edge of the wavezone, respectively.

The complicated form of the functions $F(r)$ and $U(r)$ in Eq.~\eqref{eq:SN} makes solving for the recursion relations challenging, even with the assistance of computer algebra packages. Such recursion relations are not unique but the ones we identify have 13 and 14 terms for the infinity and horizon expansions, respectively. The recursion relations we compute are too lengthy to be displayed here but we make them available in electronic format online \cite{Warburton:website}. To check the recursion relations we set $a^{\infty/H}_0 = 1$ (we are free to set this to any non-zero number as we are solving for the homogeneous solutions), compute a number of terms in the expansions, substitute the resulting expansion back into the homogeneous field equation \eqref{eq:SN} and check that is it satisfied. For the 2-2 mode with the particle on the ISCO and $a\lesssim0.99M$ we find we can satisfy the field equation at $r_\text{out}=100M$ and $r_\text{in} = r_+ + 10^{-2}M$ to over 100 significant digits with ease. As we increase the spin of the black hole the outer boundary doesn't need to move but we find we must move the inner boundary radius inwards so that by the time we reach $\epsilon=10^{-13}$ we must place the inner boundary at $r_\text{in} = r_+ + 10^{-15}M$ to achieve similar accuracy.

Lastly, we briefly mention an improvement in the practical application of the variations of parameters method used to construct the inhomogeneous solutions by integrating the homogeneous solutions against the source. With this method it is necessary to compute the Wronskian of homogeneous solutions and in Ref.~\cite{Hughes:1999bq} this is calculated at a large radius. To achieve this a variant of Richardson extrapolation was used to accurately calculate the `inner' homogeneous solution at large radii. This step is unnecessary as the Wronskian, defined with derivatives with respect to $r_*$, is a constant for all $r$ and so can be calculated at any suitable radius. We find it convenient to calculate the Wronskian at the particle's orbital radius where we can calculate the homogeneous solutions to high accuracy via direct numerical integration of the field equation.

As a test of our new Sasaki-Nakamura code we compared our results for the fluxes against those of Ref.~\cite{Hughes:1999bq} for $a\le0.99M$, finding agreement to over $8$ significant figures, which is consistent with the given error bars.  We also compare against results of Ref.~\cite{Throwe:in_prep}, which solves the Teukolsky equation as a series of special functions \cite{Sasaki:1981sx}, finding $13$ significant figures of agreement.

\subsection{Comparison of results}\label{sec:results_comparison}

\begin{table*}
\begin{center}
\begin{tabular}{c | c  c | c c }
\hline\hline
$\epsilon$ & $\dot{\mathcal{E}}_\infty$ & $\dot{\mathcal{E}}_H$ & $\Delta_\text{rel} \dot{\mathcal{E}}_\infty$ & $\Delta_\text{rel} \dot{\mathcal{E}}_H$ \\
\hline
$10^{-1}$	& $1.71745312\times 10^{-2}$	& $-3.11049626\times 10^{-3}$	& $3.5\times 10^{-1}$	& $6.0\times 10^{-1}$	\\
$10^{-2}$	& $5.33839225\times 10^{-3}$	& $-1.43536969\times 10^{-3}$	& $6.6\times 10^{-2}$	& $1.5\times 10^{-1}$	\\
$10^{-3}$	& $1.21422396\times 10^{-3}$	& $-3.50339388\times 10^{-4}$	& $1.3\times 10^{-2}$	& $3.2\times 10^{-2}$	\\
$10^{-4}$	& $2.64399402\times 10^{-4}$	& $-7.74081267\times 10^{-5}$	& $2.9\times 10^{-3}$	& $6.8\times 10^{-3}$	\\
$10^{-5}$	& $5.70914114\times 10^{-5}$	& $-1.67667887\times 10^{-5}$	& $6.1\times 10^{-4}$	& $1.5\times 10^{-3}$	\\
$10^{-6}$	& $1.23059093\times 10^{-5}$	& $-3.61646072\times 10^{-6}$	& $1.3\times 10^{-4}$	& $3.2\times 10^{-4}$	\\
$10^{-7}$	& $2.65150343\times 10^{-6}$	& $-7.79336251\times 10^{-7}$	& $2.9\times 10^{-5}$	& $6.8\times 10^{-5}$	\\
$10^{-8}$	& $5.71261941\times 10^{-7}$	& $-1.67911892\times 10^{-7}$	& $6.1\times 10^{-6}$	& $1.5\times 10^{-5}$	\\
$10^{-9}$	& $1.23075261\times 10^{-7}$	& $-3.61759396\times 10^{-8}$	& $1.3\times 10^{-6}$	& $3.2\times 10^{-6}$	\\
$10^{-10}$	& $2.65157915\times 10^{-8}$	& $-7.79388978\times 10^{-9}$	& $2.9\times 10^{-7}$	& $6.8\times 10^{-7}$	\\
$10^{-11}$	& $5.71265599\times 10^{-9}$	& $-1.67914365\times 10^{-9}$	& $6.1\times 10^{-8}$	& $1.5\times 10^{-7}$	\\
$10^{-12}$	& $1.23075461\times 10^{-9}$	& $-3.61760597\times 10^{-10}$	& $1.3\times 10^{-8}$	& $3.2\times 10^{-8}$	\\
$10^{-13}$	& $2.65158073\times 10^{-10}$	& $-7.79389645\times 10^{-11}$	& $2.9\times 10^{-9}$	& $6.8\times 10^{-9}$	\\
\hline
\end{tabular}
\end{center}
\caption{Sample numerical results and their comparison with the analytic formula for the gravitational 2-2 mode. The second and third columns give the flux radiated to infinity and through the horizon, respectively. The fourth and fifth columns give the relative difference between the numerical results and the analytic formulae, i.e., $\Delta_\text{rel}\dot{\mathcal{E}}_{\infty/H} = 1-\dot{\mathcal{E}}^\text{(num)}_{\infty/H}/\dot{\mathcal{E}}^\text{(an)}_{\infty/H}$. Note that the data in columns two and three has been adimensionalized so that $\dot{\mathcal{E}}^\text{(here)}\equiv (M/m_0)^2\dot{\mathcal{E}}$.}\label{table:results}
\end{table*}

In this section we compare the fluxes computed with our new numerical codes against our analytic flux formulae given by Eqs.~\eqref{GloriousHorizonScalar} and \eqref{GloriousInfinityScalar} for the scalar case and Eqs.~\eqref{GloriousMagnificentHorizonGravity} and \eqref{GloriousMagnificentInfinityGravity} for the gravitational case. The results for the scalar and gravitational cases are very similar and so we will concentrate on the physically more interesting gravitational case.

In making our calculations we need to evaluate the spin-weighted spheroidal-harmonics and their eigenvalues which are used in both the analytic formula and the numerical procedure. For the spin-0 harmonics we use Mathematica's inbuilt \texttt{SpheroidalPS} and \texttt{SpheroidalEigenvalue} functions. For the spin-2 harmonics we use the spectral decomposition method described in Appendix A of Ref.~\cite{Hughes:1999bq}. For ease of comparison we provide a Mathematica notebook online that evaluates the analytic flux formulae \cite{Warburton:website}.

Our main results are presented in Figs.~\ref{fig:infFlux}, \ref{fig:oscillations} and Table \ref{table:results} for the gravitational case. For the scalar case we give numerical data in Table \ref{table:results_scalar}. For small $\epsilon$ we find the numerical and analytic results agree, as expected. As an example, for the gravitational 2-2 mode we find, for the flux at infinity, that the relative difference between the analytically calculated flux, $\dot{\mathcal{E}}^\text{(an)}_\infty$, and numerically calculated flux, $\dot{\mathcal{E}}^\text{(num)}_\infty$, is around 6.6\% for $\epsilon=10^{-2}$. The agreement improves by $\epsilon=10^{-13}$ to over 8 significant figures.

For all modes the horizon flux scales as $\epsilon^{2/3}$. On the other hand, the scaling of the energy flux radiated to infinity depends on the mode in question, going as $\epsilon^p$ where $p=4/3\text{Re}[h]$. For modes with $m\sim\ell$ the scaling exponent is $p=2/3$ but for low $m$ modes $p$ is larger. Which modes are dominant or subdominant is illustrated in Fig.~\ref{fig:modes} for $\ell\le15$. For $\ell=2$ this difference in scaling can be seen explicitly by comparing the two plots in Fig.~\ref{fig:infFlux}. In addition to the leading order scaling, the horizon flux and the infinity flux for modes with $p=2/3$ exhibit oscillations. Taking the ratio of the horizon and infinity fluxes removes this leading-order behavior and makes the oscillations clear, as we show in Fig.~\ref{fig:oscillations}.

The excellent agreement we observe between our analytical and numerical results gives us confidence in both. In particular, numerical codes often struggle in such high-spin regimes and we envisage that our analytic formula will provide a valuable benchmark for future numerical work on rapidly rotating black holes.

\begin{table*}
\begin{center}
\begin{tabular}{c | c  c | c c }
\hline\hline
$\epsilon$ & $\dot{\mathcal{E}}_\infty$ & $\dot{\mathcal{E}}_H$ & $\Delta_\text{rel} \dot{\mathcal{E}}_\infty$ & $\Delta_\text{rel} \dot{\mathcal{E}}_H$ \\
\hline
$10^{-1}$	& $5.85189833\times 10^{-4}$	& $-3.34966656\times 10^{-4}$	& $-7.4\times 10^{-1}$	& $6.1\times 10^{-1}$	\\
$10^{-2}$	& $1.06917341\times 10^{-4}$	& $-1.56686813\times 10^{-4}$	& $-4.6\times 10^{-1}$	& $1.5\times 10^{-1}$	\\
$10^{-3}$	& $1.74509810\times 10^{-5}$	& $-3.84548334\times 10^{-5}$	& $-1.2\times 10^{-1}$	& $3.3\times 10^{-2}$	\\
$10^{-4}$	& $3.48448909\times 10^{-6}$	& $-8.51021948\times 10^{-6}$	& $-2.6\times 10^{-2}$	& $7.2\times 10^{-3}$	\\
$10^{-5}$	& $7.30812181\times 10^{-7}$	& $-1.84356886\times 10^{-6}$	& $-5.6\times 10^{-3}$	& $1.6\times 10^{-3}$	\\
$10^{-6}$	& $1.57583045\times 10^{-7}$	& $-3.97754374\times 10^{-7}$	& $-1.2\times 10^{-3}$	& $3.3\times 10^{-4}$	\\
$10^{-7}$	& $3.38175817\times 10^{-8}$	& $-8.56962339\times 10^{-8}$	& $-2.6\times 10^{-4}$	& $7.2\times 10^{-5}$	\\
$10^{-8}$	& $7.28828188\times 10^{-9}$	& $-1.84681469\times 10^{-8}$	& $-5.5\times 10^{-5}$	& $1.6\times 10^{-5}$	\\
$10^{-9}$	& $1.57315673\times 10^{-9}$	& $-3.97796344\times 10^{-9}$	& $-1.2\times 10^{-5}$	& $3.3\times 10^{-6}$	\\
$10^{-10}$	& $3.37478969\times 10^{-10}$	& $-8.57211607\times 10^{-10}$	& $-2.6\times 10^{-6}$	& $7.2\times 10^{-7}$	\\
$10^{-11}$	& $7.31738299\times 10^{-11}$	& $-1.84647008\times 10^{-10}$	& $-5.6\times 10^{-7}$	& $1.6\times 10^{-7}$	\\
$10^{-12}$	& $1.56369119\times 10^{-11}$	& $-3.97866228\times 10^{-11}$	& $-1.2\times 10^{-7}$	& $3.4\times 10^{-8}$	\\
$10^{-13}$	& $3.40087145\times 10^{-12}$	& $-8.57097932\times 10^{-12}$	& $-2.6\times 10^{-8}$	& $7.2\times 10^{-9}$	\\
\hline
\end{tabular}
\end{center}
\caption{The same as Table \ref{table:results} but for a particle carrying scalar charge orbiting at the ISCO. The data in columns two and three has been adimensionalized so that $\dot{\mathcal{E}}^\text{(here)}\equiv (M/q)^2\dot{\mathcal{E}}$.}\label{table:results_scalar}
\end{table*}

\acknowledgements{We thank Alex Lupsasca and Andy Strominger for helpful discussions.  We also thank Scott Hughes for providing sample numerical flux data for comparison in the gravitational case and Leor Barack for comments on a draft of this work. S.G. and A.P. were supported in part by NSF grant 1205550. N.W. gratefully acknowledges support from a Marie Curie International Outgoing Fellowship (PIOF-GA-2012-627781).

\appendix

\section{Near-Horizon Limits and Symmetries}\label{app:limits}

In this appendix we review the NHEK limits and how their enhanced symmetry naturally assigns a conformal weight $h$ to certain solutions of the wave equation.  While all of this material has appeared in some form in the literature, the references vary in their choices of notation, coordinate patch, and symmetry algebra basis.  We present the relevant results here with choices suited to our calculation.

\subsection{Far limit}

A convenient form for the Kerr exterior metric in BL coordinates is
\begin{align}\label{BLKerr}
ds^2 & = - \frac{\Delta}{\rho^2} \left(dt-a \sin^2 \! \theta d\phi \right)^2 + \frac{\sin^2 \! \theta}{\rho^2}\left((r^2+a^2) d\phi - a dt \right)^2 \nonumber \\ & \qquad +\frac{\rho^2}{\Delta} dr^2 + \rho^2 d\theta^2,
\end{align}
where $\Delta=r^2-2Mr +a^2$ and $\rho^2=r^2+a^2 \cos^2\theta$.  Setting $a=M$ (equivalently $\epsilon=0$) gives extremal Kerr.  More formally, we could introduce an auxiliary parameter $\delta$ by
\begin{align}\label{epsilonbar}
\epsilon=\sqrt{1-(a/M)^2}=\bar{\epsilon} \delta
\end{align}
for some fixed $\bar{\epsilon}$.  Letting $\delta \rightarrow 0$ at fixed BL coordinates produces extremal Kerr.  In the language of Geroch \cite{geroch1969}, we use the BL coordinates to identify metrics at different values of $\delta$.

\subsection{Near-horizon limit}\label{app:near-horizon}

We are free, however, to identify the metrics differently.  If we still use \eqref{epsilonbar} but instead hold fixed
\begin{align}\label{scaling}
\bar{x}=\frac{x}{\delta}=\frac{r-r_+}{r_+}\delta^{-1} , \quad \bar{t}=\frac{t}{2M}\delta, \quad \bar{\phi} = \phi - \frac{t}{2M},
\end{align}
then letting $\delta \rightarrow 0$ gives
\begin{align}\label{near-NHEK}
ds^2 & = 2M^2 \Gamma(\theta)\Bigg\{-\bar{x}(\bar{x}+2\bar{\epsilon}) d\bar{t}^2 + \frac{d\bar{x}^2}{\bar{x}(\bar{x}+2\bar{\epsilon})}  \nonumber \\ & \qquad + d\theta^2 + \Lambda^2(\theta) [d\bar{\phi} + (\bar{x}+\bar{\epsilon})d\bar{t}]^2 \Bigg\},
\end{align}
where $\Gamma(\theta)=(1+\cos^2\!\theta)/2$ and $\Lambda(\theta)=2\sin \theta/(1+\cos^2 \! \theta)$.  This is the ``near-NHEK'' metric \cite{bredbergetal2010}. This limit is expected to be useful for near-extremal, near-horizon physics.  It corresponds to the lowest ``crack'' in the throats diagram, Fig.~\ref{fig:throats}.

It is also the near-horizon metric in the more pedestrian sense that it agrees with near-extremal Kerr near the horizon. That is to say, Eq.~\eqref{near-NHEK} may also be obtained by using the coordinates \eqref{scaling} with $\delta=1$ and using $x \ll 1$ in the metric components, keeping to leading order in each component.  The redefinition of the $\phi$ coordinate in \eqref{scaling} is essential for the resulting metric to be non-singular, making these ``good'' near-horizon coordinates.

Consider a scalar field $\Phi$ in non-extremal Kerr ($\epsilon >0$) with the usual harmonic $t$ and $\phi$ dependence.  Expressing in the scaled coordinates \eqref{scaling} gives
\begin{align}\label{nearmode}
\Phi \sim e^{-i \omega t} e^{i m \phi} = e^{- i \bar{\omega} \bar{t}} e^{i m \bar{\phi}},
\end{align}
where we define
\begin{align}
\bar{\omega} = \frac{2M \omega - m}{\delta}.
\end{align}
If we have a family of scalar fields, one for each $\epsilon$, then for this family to have a good near-horizon limit $\omega$ must approach $m/(2M)$ linearly with $\epsilon$. (Recall that $1/(2M)$ is the extremal limit of the horizon angular velocity.)  For a circular orbit of angular velocity $\Omega$ we have $\omega = m \Omega$.  Thus for the associated mode functions to have a good near-horizon limit, $\Omega$ must approach the extremal horizon frequency linearly with $\epsilon$.

For the ISCO, $\Omega -1/(2M) \sim \epsilon^{2/3}$ and hence $\bar{\omega} \sim \delta^{-1/3}$.  The $n$ defined in the text \eqref{sigman} corresponds to $\bar{\omega}/\bar{\epsilon}$ with $\delta=1$, explaining $n \sim \epsilon^{-1/3}$ as $\epsilon \rightarrow 0$ \eqref{n2}.  The coordinate position of the ISCO also diverges in this limit, since $x_0 \sim \epsilon^{2/3}$  and hence $\bar{x} \sim \delta^{-1/3}$.  Thus from the near-NHEK point of view, the ISCO orbits infinitely far away and infinitely fast.  This is the physical origin of the infinitely oscillating phases and the need for large-$n$ asymptotics.

\subsection{Intermediate (ISCO) limit}\label{app:intermediate}

In order to avoid these difficulties one could instead define an alternate limit by keeping \eqref{epsilonbar} but replacing \eqref{scaling} with
\begin{align}\label{scaling2}
\tilde{x}=\frac{r-r_+}{r_+}\delta^{-2/3} , \quad \tilde{t}=\frac{t}{2M}\delta^{2/3}, \quad \tilde{\phi} = \phi - \frac{t}{2M},
\end{align}
which will keep the ISCO radius and frequency finite.  In this limit the metric becomes
\begin{align}\label{NHEK}
ds^2 & = 2M^2 \Gamma \Bigg\{-\tilde{x}^2 d\tilde{t}^2 + \frac{d\tilde{x}^2}{\tilde{x}^2}  + d\theta^2 + \Lambda^2 [d\tilde{\phi} + \tilde{x}d\tilde{t}]^2 \Bigg\},
\end{align}
where $\Gamma$ and $\Lambda$ are given below \eqref{near-NHEK}. This metric agrees with the near metric \eqref{near-NHEK} when $\bar{x} \gg \bar{\epsilon}$ and with the far metric (extremal Kerr) when $x \ll 1$. It too can be derived the pedestrian way, that is to say, Eq.~\eqref{NHEK} may also be obtained by using the coordinates \eqref{scaling} with $\delta=1$ and using $\epsilon\ll x \ll 1$ in the metric components, keeping to leading order in each component.
The ISCO limit is thus intermediate between the near and far regions, corresponding to the middle region in the throats diagram, Fig.~\ref{fig:throats}.

Eq.~\eqref{NHEK} is a non-singular spacetime; in fact it is \textit{diffeomorphic} to \eqref{near-NHEK}.  Eq.~\eqref{NHEK} is generally called ``NHEK'' or``Poincare NHEK" and it is the form originally discovered in \cite{bardeen-horowitz1999} as a limit of precisely extremal Kerr. This metric also approximates the near-horizon region of precisely extremal black hole in the pedestrian sense. Poincare NHEK and near-NHEK cover different patches of the maximally extended spacetime.

Despite its adaptation to the ISCO, the limit \eqref{scaling2} does not appear to be useful in calculating the radiation from a particle orbiting there.   Since the metric does not agree with near-extreme Kerr at the horizon or at  infinity, it cannot be used to impose boundary conditions at either place.  We have found it more useful to include the ISCO region in our ``near'' region in the main body for the purposes of calculation, defined as $x \ll 1$ without deciding between $x \sim \epsilon$ (the near-horizon region) and $x \sim \epsilon^{2/3}$ (the intermediate region).  Note that the ISCO is \textit{not} in our region of overlap, since that region has $x \gg \epsilon^{2/3}$ (as well as $x \ll 1$).  Correspondingly, the wave equation associated with the ISCO limit, that is the NHEK wave equation, does not appear explicitly in our calculation. Note however the connection to the extremal calculation of \cite{porfyriadis-strominger2014}, discussed in the Introduction and in Section \ref{sec:agreement}, which involved solving explicitly the NHEK wave equation.

\subsection{Symmetry Group and Conformal Weights}\label{app:symmetry}

The NHEK spacetime has an enhanced $SL(2,R) \times U(1)$ isometry group.  The explicit form of the Killing fields depends on the coordinates and the choice of basis.  We will be agnostic to both, and simply name the Killing fields $H_0$, $H_\pm$ and $W_0$, demanding only the $SL(2,R) $ commutation relations $[H_0,H_{\pm}]=\mp H_{\pm}$, $[H_+,H_-]=2H_0$ and the $U(1)$ generator $W_0$ commuting with everything.

For any complex number $h$ and integer $m$, an infinite-dimensional representation $\{ \psi_{h,m,k} \}$ with $k \geq 0$ may be constructed as follows.  The member $\psi_{h,m,0}$ should satisfy the highest-weight condition,
\begin{subequations}\label{highest-weight}
\begin{align}
\mathcal{L}_{H_+} \psi_{h,m,0} & = 0 \label{0eqn} \\
\mathcal{L}_{H_0} \psi_{h,m,0} & = h \psi ,\label{heqn}
\end{align}
\end{subequations}
together with $\mathcal{L}_{W_0} \psi_{h,m,0}  = i m \psi\label{meqn}$.   The remaining members of the representation are formed by repeated application of $H_-$,
\begin{align}\label{descent}
\psi_{h,m,k} = (\mathcal{L}_{H_-})^k \psi_{h,m,0}.
\end{align}
Here $\mathcal{L}$ is the Lie derivative.  Since $SL(2,R)$ is not compact, this tower does not terminate and the representation is infinite-dimensional.

Solutions to the wave equation may be organized in representations of the isometry group.
For simplicity, we work in the case of a scalar field $\Phi$ and drop circumflexes in order to be agnostic between the coordinates \eqref{near-NHEK} and \eqref{NHEK}. Adopting the decomposition
\begin{align}
\Phi = Q(x,t) S(\theta) e^{i m \phi},
\end{align}
with $S$ satisfying the angular equation \eqref{SeqnSimp}, the wave equation implies:
\begin{align}\label{porpoise}
\left[ \mathcal{L}_{H_0}( \mathcal{L}_{H_0}-1) -  \mathcal{L}_{H_-}  \mathcal{L}_{H_+} \right]\Phi  = (K-2m^2) \Phi.
\end{align}
The operator on the LHS is the Casimir of $SL(2,R)$; thus $\Phi$ is an eigenstate of the Casimir with eigenvalue $K-2m^2$. If $\Phi$ is also highest-weight \eqref{highest-weight} then this becomes
\begin{equation}\label{dolphin}
h(h-1) +2m^2 -K=0,
\end{equation}
which is solved by
\begin{equation}\label{otter}
 h = \frac{1}{2} \pm \sqrt{K-2 m^2+\frac{1}{4}}.
\end{equation}
(In the text we define $h$ to be the plus branch of \eqref{otter}. The minus branch appears as $1-h$ in (e.g.) \eqref{Roverlap}.)  The highest-weight condition thus replaces the second-order differential equation \eqref{porpoise} with two first-order equations \eqref{0eqn} and \eqref{heqn}.   Once the highest-weight solution is found by solving these equations, further solutions arise from its descendants via \eqref{descent}.  In this way one constructs an infinite tower of solutions for each $\{K,m\}$ angular mode.  It is generally believed that these comprise a complete basis for solutions of the wave equation.  Our near equation \eqref{neareqn} is the radial near-NHEK wave equation for the ansatz $\Phi = R(x) e^{-i (n-m)\epsilon t} S(\theta) e^{i m \phi}$. Therefore, the near solutions $R^{\rm near}(x) e^{-i (n-m)\epsilon t} S(\theta) e^{i m \phi}$ must be expressible as linear combinations of the members of the highest weight representations labelled by \eqref{otter}. 

\subsection{Where is the extremal Kerr ISCO?}

Having introduced three inequivalent limits and discussed their properties, we conclude with a discussion of the ``location'' of the extremal Kerr ISCO, an innocent question with an amusingly complicated answer.

The far limit produces the extremal Kerr exterior and the ISCO exits the domain, approaching $r=M$ ($x=0$).  Correspondingly, every equatorial (prograde) circular orbit in extremal Kerr is stable.  If we complete the domain by including the horizon (e.g. taking the limit at fixed Doran coordinate), then the ISCO approaches the horizon generators \cite{jacobson2011}.

The near-horizon limit produces near-NHEK \eqref{near-NHEK} and the ISCO also exits the domain, approaching $\bar{x}_0 \rightarrow \infty$.  Correspondingly, there are no stable circular orbits in near-NHEK \cite{hadar-porfyriadis-strominger2014}.

The intermediate limit produces NHEK \eqref{NHEK} and ISCO achieves a finite coordinate value $\tilde{x}_0=2^{1/3}$.  Is this, then, the location of the ISCO?  No: redefining \eqref{scaling2} by $\tilde{x} \rightarrow c \tilde{x}$ and $\tilde{t} \rightarrow \tilde{t}/c$ for a number $c$, we produce the same limiting metric \eqref{NHEK} but find that the ISCO instead approached $\tilde{x}_0=c \times 2^{1/3}$.  We can therefore put the ISCO anywhere we want.  Within \eqref{NHEK} itself this can be seen as the fact that $\tilde{x} \rightarrow c \tilde{x}$ and $\tilde{t} \rightarrow \tilde{t}/c$ is a symmetry---the ``dilation'' member of the enhanced $SL(2,R)$ symmetry group.  This symmetry maps circular orbits to circular orbits, making all circular orbits physically equivalent within NHEK.\footnote{The position of the ISCO gains meaning only through matching to the external region (far limit), which breaks the $SL(2,R)$ symmetries.}  In particular, they are all marginally stable \cite{hadar-porfyriadis-strominger2014}, like the ISCO.

Where is the extremal Kerr ISCO?  It's on the horizon in the far limit, at infinity in the near-horizon limit, and in the intermediate limit, it is everywhere!

\nopagebreak

\bibliographystyle{apsrev4-1}
\bibliography{paper4}

\begin{thebibliography}{31}%
\makeatletter
\providecommand \@ifxundefined [1]{%
 \@ifx{#1\undefined}
}%
\providecommand \@ifnum [1]{%
 \ifnum #1\expandafter \@firstoftwo
 \else \expandafter \@secondoftwo
 \fi
}%
\providecommand \@ifx [1]{%
 \ifx #1\expandafter \@firstoftwo
 \else \expandafter \@secondoftwo
 \fi
}%
\providecommand \natexlab [1]{#1}%
\providecommand \enquote  [1]{``#1''}%
\providecommand \bibnamefont  [1]{#1}%
\providecommand \bibfnamefont [1]{#1}%
\providecommand \citenamefont [1]{#1}%
\providecommand \href@noop [0]{\@secondoftwo}%
\providecommand \href [0]{\begingroup \@sanitize@url \@href}%
\providecommand \@href[1]{\@@startlink{#1}\@@href}%
\providecommand \@@href[1]{\endgroup#1\@@endlink}%
\providecommand \@sanitize@url [0]{\catcode `\\12\catcode `\$12\catcode
  `\&12\catcode `\#12\catcode `\^12\catcode `\_12\catcode `\%12\relax}%
\providecommand \@@startlink[1]{}%
\providecommand \@@endlink[0]{}%
\providecommand \url  [0]{\begingroup\@sanitize@url \@url }%
\providecommand \@url [1]{\endgroup\@href {#1}{\urlprefix }}%
\providecommand \urlprefix  [0]{URL }%
\providecommand \Eprint [0]{\href }%
\providecommand \doibase [0]{http://dx.doi.org/}%
\providecommand \selectlanguage [0]{\@gobble}%
\providecommand \bibinfo  [0]{\@secondoftwo}%
\providecommand \bibfield  [0]{\@secondoftwo}%
\providecommand \translation [1]{[#1]}%
\providecommand \BibitemOpen [0]{}%
\providecommand \bibitemStop [0]{}%
\providecommand \bibitemNoStop [0]{.\EOS\space}%
\providecommand \EOS [0]{\spacefactor3000\relax}%
\providecommand \BibitemShut  [1]{\csname bibitem#1\endcsname}%
\let\auto@bib@innerbib\@empty
\bibitem [{\citenamefont {{Le Tiec}}(2014)}]{letiec2014}%
  \BibitemOpen
  \bibfield  {author} {\bibinfo {author} {\bibfnamefont {A.}~\bibnamefont {{Le
  Tiec}}},\ }\href {\doibase 10.1142/S0218271814300225} {\bibfield  {journal}
  {\bibinfo  {journal} {International Journal of Modern Physics D}\ }\textbf
  {\bibinfo {volume} {23}},\ \bibinfo {eid} {1430022} (\bibinfo {year}
  {2014})},\ \Eprint {http://arxiv.org/abs/1408.5505} {arXiv:1408.5505 [gr-qc]}
  \BibitemShut {NoStop}%
\bibitem [{\citenamefont {{Porfyriadis}}\ and\ \citenamefont
  {{Strominger}}(2014)}]{porfyriadis-strominger2014}%
  \BibitemOpen
  \bibfield  {author} {\bibinfo {author} {\bibfnamefont {A.~P.}\ \bibnamefont
  {{Porfyriadis}}}\ and\ \bibinfo {author} {\bibfnamefont {A.}~\bibnamefont
  {{Strominger}}},\ }\href {\doibase 10.1103/PhysRevD.90.044038} {\bibfield
  {journal} {\bibinfo  {journal} {\prd}\ }\textbf {\bibinfo {volume} {90}},\
  \bibinfo {eid} {044038} (\bibinfo {year} {2014})},\ \Eprint
  {http://arxiv.org/abs/1401.3746} {arXiv:1401.3746 [hep-th]} \BibitemShut
  {NoStop}%
\bibitem [{\citenamefont {{Hadar}}\ \emph {et~al.}(2014)\citenamefont
  {{Hadar}}, \citenamefont {{Porfyriadis}},\ and\ \citenamefont
  {{Strominger}}}]{hadar-porfyriadis-strominger2014}%
  \BibitemOpen
  \bibfield  {author} {\bibinfo {author} {\bibfnamefont {S.}~\bibnamefont
  {{Hadar}}}, \bibinfo {author} {\bibfnamefont {A.~P.}\ \bibnamefont
  {{Porfyriadis}}}, \ and\ \bibinfo {author} {\bibfnamefont {A.}~\bibnamefont
  {{Strominger}}},\ }\href {\doibase 10.1103/PhysRevD.90.064045} {\bibfield
  {journal} {\bibinfo  {journal} {\prd}\ }\textbf {\bibinfo {volume} {90}},\
  \bibinfo {eid} {064045} (\bibinfo {year} {2014})},\ \Eprint
  {http://arxiv.org/abs/1403.2797} {arXiv:1403.2797 [hep-th]} \BibitemShut
  {NoStop}%
\bibitem [{\citenamefont {{Hadar}}\ \emph {et~al.}(2015)\citenamefont
  {{Hadar}}, \citenamefont {{Porfyriadis}},\ and\ \citenamefont
  {{Strominger}}}]{hadar-porfyriadis-strominger2015}%
  \BibitemOpen
  \bibfield  {author} {\bibinfo {author} {\bibfnamefont {S.}~\bibnamefont
  {{Hadar}}}, \bibinfo {author} {\bibfnamefont {A.~P.}\ \bibnamefont
  {{Porfyriadis}}}, \ and\ \bibinfo {author} {\bibfnamefont {A.}~\bibnamefont
  {{Strominger}}},\ }\href@noop {} {\bibfield  {journal} {\bibinfo  {journal}
  {ArXiv e-prints}\ } (\bibinfo {year} {2015})},\ \Eprint
  {http://arxiv.org/abs/1504.07650} {arXiv:1504.07650 [hep-th]} \BibitemShut
  {NoStop}%
\bibitem [{\citenamefont {{Bardeen}}\ and\ \citenamefont
  {{Horowitz}}(1999)}]{bardeen-horowitz1999}%
  \BibitemOpen
  \bibfield  {author} {\bibinfo {author} {\bibfnamefont {J.}~\bibnamefont
  {{Bardeen}}}\ and\ \bibinfo {author} {\bibfnamefont {G.~T.}\ \bibnamefont
  {{Horowitz}}},\ }\href {\doibase 10.1103/PhysRevD.60.104030} {\bibfield
  {journal} {\bibinfo  {journal} {\prd}\ }\textbf {\bibinfo {volume} {60}},\
  \bibinfo {eid} {104030} (\bibinfo {year} {1999})},\ \Eprint
  {http://arxiv.org/abs/hep-th/9905099} {hep-th/9905099} \BibitemShut {NoStop}%
\bibitem [{\citenamefont {{Guica}}\ \emph {et~al.}(2009)\citenamefont
  {{Guica}}, \citenamefont {{Hartman}}, \citenamefont {{Song}},\ and\
  \citenamefont {{Strominger}}}]{kerrCFT}%
  \BibitemOpen
  \bibfield  {author} {\bibinfo {author} {\bibfnamefont {M.}~\bibnamefont
  {{Guica}}}, \bibinfo {author} {\bibfnamefont {T.}~\bibnamefont {{Hartman}}},
  \bibinfo {author} {\bibfnamefont {W.}~\bibnamefont {{Song}}}, \ and\ \bibinfo
  {author} {\bibfnamefont {A.}~\bibnamefont {{Strominger}}},\ }\href {\doibase
  10.1103/PhysRevD.80.124008} {\bibfield  {journal} {\bibinfo  {journal}
  {\prd}\ }\textbf {\bibinfo {volume} {80}},\ \bibinfo {eid} {124008} (\bibinfo
  {year} {2009})},\ \Eprint {http://arxiv.org/abs/0809.4266} {arXiv:0809.4266
  [hep-th]} \BibitemShut {NoStop}%
\bibitem [{\citenamefont {{Buonanno}}\ and\ \citenamefont
  {{Damour}}(1999)}]{buonanno-damour1999}%
  \BibitemOpen
  \bibfield  {author} {\bibinfo {author} {\bibfnamefont {A.}~\bibnamefont
  {{Buonanno}}}\ and\ \bibinfo {author} {\bibfnamefont {T.}~\bibnamefont
  {{Damour}}},\ }\href {\doibase 10.1103/PhysRevD.59.084006} {\bibfield
  {journal} {\bibinfo  {journal} {\prd}\ }\textbf {\bibinfo {volume} {59}},\
  \bibinfo {eid} {084006} (\bibinfo {year} {1999})},\ \Eprint
  {http://arxiv.org/abs/gr-qc/9811091} {gr-qc/9811091} \BibitemShut {NoStop}%
\bibitem [{\citenamefont {{Hubeny}}(1999)}]{hubeny1999}%
  \BibitemOpen
  \bibfield  {author} {\bibinfo {author} {\bibfnamefont {V.~E.}\ \bibnamefont
  {{Hubeny}}},\ }\href {\doibase 10.1103/PhysRevD.59.064013} {\bibfield
  {journal} {\bibinfo  {journal} {\prd}\ }\textbf {\bibinfo {volume} {59}},\
  \bibinfo {eid} {064013} (\bibinfo {year} {1999})},\ \Eprint
  {http://arxiv.org/abs/gr-qc/9808043} {gr-qc/9808043} \BibitemShut {NoStop}%
\bibitem [{\citenamefont {{Jacobson}}\ and\ \citenamefont
  {{Sotiriou}}(2009)}]{jacobson-sotiriou2009}%
  \BibitemOpen
  \bibfield  {author} {\bibinfo {author} {\bibfnamefont {T.}~\bibnamefont
  {{Jacobson}}}\ and\ \bibinfo {author} {\bibfnamefont {T.~P.}\ \bibnamefont
  {{Sotiriou}}},\ }\href {\doibase 10.1103/PhysRevLett.103.141101} {\bibfield
  {journal} {\bibinfo  {journal} {Physical Review Letters}\ }\textbf {\bibinfo
  {volume} {103}},\ \bibinfo {eid} {141101} (\bibinfo {year} {2009})},\ \Eprint
  {http://arxiv.org/abs/0907.4146} {arXiv:0907.4146 [gr-qc]} \BibitemShut
  {NoStop}%
\bibitem [{\citenamefont {{Ori}}\ and\ \citenamefont
  {{Thorne}}(2000)}]{ori-thorne2000}%
  \BibitemOpen
  \bibfield  {author} {\bibinfo {author} {\bibfnamefont {A.}~\bibnamefont
  {{Ori}}}\ and\ \bibinfo {author} {\bibfnamefont {K.~S.}\ \bibnamefont
  {{Thorne}}},\ }\href {\doibase 10.1103/PhysRevD.62.124022} {\bibfield
  {journal} {\bibinfo  {journal} {\prd}\ }\textbf {\bibinfo {volume} {62}},\
  \bibinfo {eid} {124022} (\bibinfo {year} {2000})},\ \Eprint
  {http://arxiv.org/abs/gr-qc/0003032} {gr-qc/0003032} \BibitemShut {NoStop}%
\bibitem [{\citenamefont {{Kesden}}(2011)}]{kesden2011}%
  \BibitemOpen
  \bibfield  {author} {\bibinfo {author} {\bibfnamefont {M.}~\bibnamefont
  {{Kesden}}},\ }\href {\doibase 10.1103/PhysRevD.83.104011} {\bibfield
  {journal} {\bibinfo  {journal} {\prd}\ }\textbf {\bibinfo {volume} {83}},\
  \bibinfo {eid} {104011} (\bibinfo {year} {2011})},\ \Eprint
  {http://arxiv.org/abs/1101.3749} {arXiv:1101.3749 [gr-qc]} \BibitemShut
  {NoStop}%
\bibitem [{\citenamefont {Barausse}\ \emph {et~al.}(2010)\citenamefont
  {Barausse}, \citenamefont {Cardoso},\ and\ \citenamefont
  {Khanna}}]{Barausse:2010ka}%
  \BibitemOpen
  \bibfield  {author} {\bibinfo {author} {\bibfnamefont {E.}~\bibnamefont
  {Barausse}}, \bibinfo {author} {\bibfnamefont {V.}~\bibnamefont {Cardoso}}, \
  and\ \bibinfo {author} {\bibfnamefont {G.}~\bibnamefont {Khanna}},\ }\href
  {\doibase 10.1103/PhysRevLett.105.261102} {\bibfield  {journal} {\bibinfo
  {journal} {Phys. Rev. Lett.}\ }\textbf {\bibinfo {volume} {105}},\ \bibinfo
  {pages} {261102} (\bibinfo {year} {2010})},\ \Eprint
  {http://arxiv.org/abs/1008.5159} {arXiv:1008.5159 [gr-qc]} \BibitemShut
  {NoStop}%
\bibitem [{\citenamefont {{Barausse}}\ \emph {et~al.}(2011)\citenamefont
  {{Barausse}}, \citenamefont {{Cardoso}},\ and\ \citenamefont
  {{Khanna}}}]{barausse-cardoso-khanna2011}%
  \BibitemOpen
  \bibfield  {author} {\bibinfo {author} {\bibfnamefont {E.}~\bibnamefont
  {{Barausse}}}, \bibinfo {author} {\bibfnamefont {V.}~\bibnamefont
  {{Cardoso}}}, \ and\ \bibinfo {author} {\bibfnamefont {G.}~\bibnamefont
  {{Khanna}}},\ }\href {\doibase 10.1103/PhysRevD.84.104006} {\bibfield
  {journal} {\bibinfo  {journal} {\prd}\ }\textbf {\bibinfo {volume} {84}},\
  \bibinfo {eid} {104006} (\bibinfo {year} {2011})},\ \Eprint
  {http://arxiv.org/abs/1106.1692} {arXiv:1106.1692 [gr-qc]} \BibitemShut
  {NoStop}%
\bibitem [{\citenamefont {{Colleoni}}\ and\ \citenamefont
  {{Barack}}(2015)}]{colleoni-barack2015}%
  \BibitemOpen
  \bibfield  {author} {\bibinfo {author} {\bibfnamefont {M.}~\bibnamefont
  {{Colleoni}}}\ and\ \bibinfo {author} {\bibfnamefont {L.}~\bibnamefont
  {{Barack}}},\ }\href {\doibase 10.1103/PhysRevD.91.104024} {\bibfield
  {journal} {\bibinfo  {journal} {\prd}\ }\textbf {\bibinfo {volume} {91}},\
  \bibinfo {eid} {104024} (\bibinfo {year} {2015})},\ \Eprint
  {http://arxiv.org/abs/1501.07330} {arXiv:1501.07330 [gr-qc]} \BibitemShut
  {NoStop}%
\bibitem [{\citenamefont {{Chrzanowski}}(1976)}]{chrzanowski1976}%
  \BibitemOpen
  \bibfield  {author} {\bibinfo {author} {\bibfnamefont {P.~L.}\ \bibnamefont
  {{Chrzanowski}}},\ }\href {\doibase 10.1103/PhysRevD.13.806} {\bibfield
  {journal} {\bibinfo  {journal} {\prd}\ }\textbf {\bibinfo {volume} {13}},\
  \bibinfo {pages} {806} (\bibinfo {year} {1976})}\BibitemShut {NoStop}%
\bibitem [{\citenamefont {{Yang}}\ \emph
  {et~al.}(2013{\natexlab{a}})\citenamefont {{Yang}}, \citenamefont {{Zhang}},
  \citenamefont {{Zimmerman}}, \citenamefont {{Nichols}}, \citenamefont
  {{Berti}},\ and\ \citenamefont {{Chen}}}]{yang-etalOne2013}%
  \BibitemOpen
  \bibfield  {author} {\bibinfo {author} {\bibfnamefont {H.}~\bibnamefont
  {{Yang}}}, \bibinfo {author} {\bibfnamefont {F.}~\bibnamefont {{Zhang}}},
  \bibinfo {author} {\bibfnamefont {A.}~\bibnamefont {{Zimmerman}}}, \bibinfo
  {author} {\bibfnamefont {D.~A.}\ \bibnamefont {{Nichols}}}, \bibinfo {author}
  {\bibfnamefont {E.}~\bibnamefont {{Berti}}}, \ and\ \bibinfo {author}
  {\bibfnamefont {Y.}~\bibnamefont {{Chen}}},\ }\href {\doibase
  10.1103/PhysRevD.87.041502} {\bibfield  {journal} {\bibinfo  {journal}
  {\prd}\ }\textbf {\bibinfo {volume} {87}},\ \bibinfo {eid} {041502} (\bibinfo
  {year} {2013}{\natexlab{a}})},\ \Eprint {http://arxiv.org/abs/1212.3271}
  {arXiv:1212.3271 [gr-qc]} \BibitemShut {NoStop}%
\bibitem [{\citenamefont {{Yang}}\ \emph
  {et~al.}(2013{\natexlab{b}})\citenamefont {{Yang}}, \citenamefont
  {{Zimmerman}}, \citenamefont {{Zengino{\u g}lu}}, \citenamefont {{Zhang}},
  \citenamefont {{Berti}},\ and\ \citenamefont {{Chen}}}]{yang-etal2013}%
  \BibitemOpen
  \bibfield  {author} {\bibinfo {author} {\bibfnamefont {H.}~\bibnamefont
  {{Yang}}}, \bibinfo {author} {\bibfnamefont {A.}~\bibnamefont {{Zimmerman}}},
  \bibinfo {author} {\bibfnamefont {A.}~\bibnamefont {{Zengino{\u g}lu}}},
  \bibinfo {author} {\bibfnamefont {F.}~\bibnamefont {{Zhang}}}, \bibinfo
  {author} {\bibfnamefont {E.}~\bibnamefont {{Berti}}}, \ and\ \bibinfo
  {author} {\bibfnamefont {Y.}~\bibnamefont {{Chen}}},\ }\href {\doibase
  10.1103/PhysRevD.88.044047} {\bibfield  {journal} {\bibinfo  {journal}
  {\prd}\ }\textbf {\bibinfo {volume} {88}},\ \bibinfo {eid} {044047} (\bibinfo
  {year} {2013}{\natexlab{b}})},\ \Eprint {http://arxiv.org/abs/1307.8086}
  {arXiv:1307.8086 [gr-qc]} \BibitemShut {NoStop}%
\bibitem [{\citenamefont {{Starobinskij}}(1973)}]{Starobisky}%
  \BibitemOpen
  \bibfield  {author} {\bibinfo {author} {\bibfnamefont {A.~A.}\ \bibnamefont
  {{Starobinskij}}},\ }\href@noop {} {\bibfield  {journal} {\bibinfo  {journal}
  {Zhurnal Eksperimentalnoi i Teoreticheskoi Fiziki}\ }\textbf {\bibinfo
  {volume} {64}},\ \bibinfo {pages} {48} (\bibinfo {year} {1973})}\BibitemShut
  {NoStop}%
\bibitem [{\citenamefont {{Starobinskij}}\ and\ \citenamefont
  {{Churilov}}(1973)}]{starobisky-churilov}%
  \BibitemOpen
  \bibfield  {author} {\bibinfo {author} {\bibfnamefont {A.~A.}\ \bibnamefont
  {{Starobinskij}}}\ and\ \bibinfo {author} {\bibfnamefont {S.~M.}\
  \bibnamefont {{Churilov}}},\ }\href@noop {} {\bibfield  {journal} {\bibinfo
  {journal} {Zhurnal Eksperimentalnoi i Teoreticheskoi Fiziki}\ }\textbf
  {\bibinfo {volume} {65}},\ \bibinfo {pages} {3} (\bibinfo {year}
  {1973})}\BibitemShut {NoStop}%
\bibitem [{\citenamefont {{Teukolsky}}\ and\ \citenamefont
  {{Press}}(1974)}]{teukolsky-press1974}%
  \BibitemOpen
  \bibfield  {author} {\bibinfo {author} {\bibfnamefont {S.~A.}\ \bibnamefont
  {{Teukolsky}}}\ and\ \bibinfo {author} {\bibfnamefont {W.~H.}\ \bibnamefont
  {{Press}}},\ }\href {\doibase 10.1086/153180} {\bibfield  {journal} {\bibinfo
   {journal} {\apj}\ }\textbf {\bibinfo {volume} {193}},\ \bibinfo {pages}
  {443} (\bibinfo {year} {1974})}\BibitemShut {NoStop}%
\bibitem [{\citenamefont {{Bardeen}}\ \emph {et~al.}(1972)\citenamefont
  {{Bardeen}}, \citenamefont {{Press}},\ and\ \citenamefont
  {{Teukolsky}}}]{bardeen-press-teukolsky1972}%
  \BibitemOpen
  \bibfield  {author} {\bibinfo {author} {\bibfnamefont {J.~M.}\ \bibnamefont
  {{Bardeen}}}, \bibinfo {author} {\bibfnamefont {W.~H.}\ \bibnamefont
  {{Press}}}, \ and\ \bibinfo {author} {\bibfnamefont {S.~A.}\ \bibnamefont
  {{Teukolsky}}},\ }\href {\doibase 10.1086/151796} {\bibfield  {journal}
  {\bibinfo  {journal} {\apj}\ }\textbf {\bibinfo {volume} {178}},\ \bibinfo
  {pages} {347} (\bibinfo {year} {1972})}\BibitemShut {NoStop}%
\bibitem [{\citenamefont {{Teukolsky}}(1973)}]{teukolsky1973}%
  \BibitemOpen
  \bibfield  {author} {\bibinfo {author} {\bibfnamefont {S.~A.}\ \bibnamefont
  {{Teukolsky}}},\ }\href {\doibase 10.1086/152444} {\bibfield  {journal}
  {\bibinfo  {journal} {\apj}\ }\textbf {\bibinfo {volume} {185}},\ \bibinfo
  {pages} {635} (\bibinfo {year} {1973})}\BibitemShut {NoStop}%
\bibitem [{\citenamefont {Warburton}\ and\ \citenamefont
  {Barack}(2010)}]{Warburton:2010eq}%
  \BibitemOpen
  \bibfield  {author} {\bibinfo {author} {\bibfnamefont {N.}~\bibnamefont
  {Warburton}}\ and\ \bibinfo {author} {\bibfnamefont {L.}~\bibnamefont
  {Barack}},\ }\href {\doibase 10.1103/PhysRevD.81.084039} {\bibfield
  {journal} {\bibinfo  {journal} {Phys.Rev.}\ }\textbf {\bibinfo {volume}
  {D81}},\ \bibinfo {pages} {084039} (\bibinfo {year} {2010})},\ \Eprint
  {http://arxiv.org/abs/1003.1860} {arXiv:1003.1860 [gr-qc]} \BibitemShut
  {NoStop}%
\bibitem [{\citenamefont {Hughes}(2000)}]{Hughes:1999bq}%
  \BibitemOpen
  \bibfield  {author} {\bibinfo {author} {\bibfnamefont {S.~A.}\ \bibnamefont
  {Hughes}},\ }\href {\doibase 10.1103/PhysRevD.65.069902,
  10.1103/PhysRevD.90.109904, 10.1103/PhysRevD.61.084004,
  10.1103/PhysRevD.63.049902, 10.1103/PhysRevD.67.089901} {\bibfield  {journal}
  {\bibinfo  {journal} {Phys.Rev.}\ }\textbf {\bibinfo {volume} {D61}},\
  \bibinfo {pages} {084004} (\bibinfo {year} {2000})},\ \Eprint
  {http://arxiv.org/abs/gr-qc/9910091} {arXiv:gr-qc/9910091 [gr-qc]}
  \BibitemShut {NoStop}%
\bibitem [{\citenamefont {{Warburton}}()}]{Warburton:website}%
  \BibitemOpen
  \bibfield  {author} {\bibinfo {author} {\bibfnamefont {N.}~\bibnamefont
  {{Warburton}}},\ }\href@noop {} {}\bibinfo {howpublished}
  {\url{http://www.nielswarburton.net}}\BibitemShut {NoStop}%
\bibitem [{\citenamefont {Sasaki}\ and\ \citenamefont
  {Nakamura}(1982)}]{Sasaki:1981sx}%
  \BibitemOpen
  \bibfield  {author} {\bibinfo {author} {\bibfnamefont {M.}~\bibnamefont
  {Sasaki}}\ and\ \bibinfo {author} {\bibfnamefont {T.}~\bibnamefont
  {Nakamura}},\ }\href {\doibase 10.1143/PTP.67.1788} {\bibfield  {journal}
  {\bibinfo  {journal} {Prog.Theor.Phys.}\ }\textbf {\bibinfo {volume} {67}},\
  \bibinfo {pages} {1788} (\bibinfo {year} {1982})}\BibitemShut {NoStop}%
\bibitem [{\citenamefont {Shah}\ \emph {et~al.}(2012)\citenamefont {Shah},
  \citenamefont {Friedman},\ and\ \citenamefont {Keidl}}]{Shah:2012gu}%
  \BibitemOpen
  \bibfield  {author} {\bibinfo {author} {\bibfnamefont {A.~G.}\ \bibnamefont
  {Shah}}, \bibinfo {author} {\bibfnamefont {J.~L.}\ \bibnamefont {Friedman}},
  \ and\ \bibinfo {author} {\bibfnamefont {T.~S.}\ \bibnamefont {Keidl}},\
  }\href {\doibase 10.1103/PhysRevD.86.084059} {\bibfield  {journal} {\bibinfo
  {journal} {Phys.Rev.}\ }\textbf {\bibinfo {volume} {D86}},\ \bibinfo {pages}
  {084059} (\bibinfo {year} {2012})},\ \Eprint {http://arxiv.org/abs/1207.5595}
  {arXiv:1207.5595 [gr-qc]} \BibitemShut {NoStop}%
\bibitem [{\citenamefont {{Throwe}}\ \emph {et~al.}()\citenamefont {{Throwe}},
  \citenamefont {{Hughes}},\ and\ \citenamefont {{Drasco}}}]{Throwe:in_prep}%
  \BibitemOpen
  \bibfield  {author} {\bibinfo {author} {\bibfnamefont {W.}~\bibnamefont
  {{Throwe}}}, \bibinfo {author} {\bibfnamefont {S.~A.}\ \bibnamefont
  {{Hughes}}}, \ and\ \bibinfo {author} {\bibfnamefont {S.}~\bibnamefont
  {{Drasco}}},\ }\href@noop {} {}\bibinfo {note} {{i}n preparation}\BibitemShut
  {NoStop}%
\bibitem [{\citenamefont {{Geroch}}(1969)}]{geroch1969}%
  \BibitemOpen
  \bibfield  {author} {\bibinfo {author} {\bibfnamefont {R.}~\bibnamefont
  {{Geroch}}},\ }\href {\doibase 10.1007/BF01645486} {\bibfield  {journal}
  {\bibinfo  {journal} {Communications in Mathematical Physics}\ }\textbf
  {\bibinfo {volume} {13}},\ \bibinfo {pages} {180} (\bibinfo {year}
  {1969})}\BibitemShut {NoStop}%
\bibitem [{\citenamefont {{Bredberg}}\ \emph {et~al.}(2010)\citenamefont
  {{Bredberg}}, \citenamefont {{Hartman}}, \citenamefont {{Song}},\ and\
  \citenamefont {{Strominger}}}]{bredbergetal2010}%
  \BibitemOpen
  \bibfield  {author} {\bibinfo {author} {\bibfnamefont {I.}~\bibnamefont
  {{Bredberg}}}, \bibinfo {author} {\bibfnamefont {T.}~\bibnamefont
  {{Hartman}}}, \bibinfo {author} {\bibfnamefont {W.}~\bibnamefont {{Song}}}, \
  and\ \bibinfo {author} {\bibfnamefont {A.}~\bibnamefont {{Strominger}}},\
  }\href {\doibase 10.1007/JHEP04(2010)019} {\bibfield  {journal} {\bibinfo
  {journal} {Journal of High Energy Physics}\ }\textbf {\bibinfo {volume}
  {4}},\ \bibinfo {eid} {19} (\bibinfo {year} {2010})},\ \Eprint
  {http://arxiv.org/abs/0907.3477} {arXiv:0907.3477 [hep-th]} \BibitemShut
  {NoStop}%
\bibitem [{\citenamefont {{Jacobson}}(2011)}]{jacobson2011}%
  \BibitemOpen
  \bibfield  {author} {\bibinfo {author} {\bibfnamefont {T.}~\bibnamefont
  {{Jacobson}}},\ }\href {\doibase 10.1088/0264-9381/28/18/187001} {\bibfield
  {journal} {\bibinfo  {journal} {Classical and Quantum Gravity}\ }\textbf
  {\bibinfo {volume} {28}},\ \bibinfo {eid} {187001} (\bibinfo {year}
  {2011})},\ \Eprint {http://arxiv.org/abs/1107.5081} {arXiv:1107.5081 [gr-qc]}
  \BibitemShut {NoStop}%
\end{thebibliography}%

\end{document}